\documentclass[preprint,12pt]{elsarticle}

\usepackage{amssymb}
\usepackage{amsmath}
\usepackage{graphicx}
\usepackage{nomencl,etoolbox,ragged2e,siunitx,mathtools}
\usepackage{setspace}
\usepackage{booktabs}
\usepackage{lscape}
\usepackage{arydshln}
\usepackage[algoruled,boxed,lined]{algorithm2e}
\usepackage{algorithmic}
\usepackage{color}
\usepackage{multirow}
\usepackage{tablefootnote}
\usepackage{physics}
\usepackage{newtxtext,newtxmath}
\usepackage{verbatim}
\journal{International Journal of Multiphase Flow}

\begin{document}
	
	\begin{frontmatter}
		
		\title{Numerical modeling of bubble--particle interaction in a volume-of-fluid framework}
		
		\author[]{Hojun Moon}
		\author[]{Jeongbo Shim, and Donghyun You\footnote{Corresponding author. E-mail: dhyou@postech.ac.kr; Phone: +82-54-279-2191; Fax: +82-54-279-3199}}
		\address{Department of Mechanical Engineering, Pohang University of Science and Technology, 77 Cheongam-ro, Nam-gu, Pohang, Gyeongbuk 37673, Republic of Korea}
		
		\begin{abstract}
		A numerical method is presented to simulate gas--liquid--solid flows with bubble--particle interaction, including particle collision, sliding, and attachment. Gas--liquid flows are simulated in an Eulerian framework using a volume-of-fluid method. Particle motions are predicted in a Lagrangian framework. Algorithms that are used to detect collision and determine the sliding or attachment of the particle are developed.  An effective bubble is introduced to model these bubble--particle interaction. The proposed numerical method is validated through experimental cases that entail the rising of a single bubble with particles. Collision and attachment probabilities obtained from the simulation are compared to model and experimental results based on bubble diameters, particle diameters, and contact angles. The particle trajectories near the bubble are presented to show differences with and without the proposed bubble--particle interaction model. The sliding and attachment of the colliding particle are observed using this model.
			
		\end{abstract}
		
		\begin{keyword}
                Bubble--particle interaction \sep
			Volume-of-fluid \sep
			Gas--liquid--solid flows \sep
			Attachment probability
		\end{keyword}
		
	\end{frontmatter}
	


	\section{Introduction}\label{sec:introduction}

	Bubble--particle interaction is a common physical phenomenon that can be observed in various industrial applications, such as in the recovery of coal and minerals from ores, de-inking of paper for the recycling of waste paper, and inclusion removal in steel industries~\cite{Nguyen2003}. Specifically, an inclusion removal refers to a technique that is widely used to manufacture ultra-clean steel. An inclusion refers to a micro-sized particle created from the oxidation of molten steel, which degrades steel quality. To enhance steel quality, the removal technique utilizes bubbles to collect inclusions by attaching them on the bubble surface~\cite{Zhang2000}. Thus, understanding the physics of inclusion attachment is directly related to collision efficiency. However, inclusion removal is a complex phenomenon that is challenging to be described by a numerical simulation. Therefore, a simple method that reflects the physical phenomena of the removal process must be used.

	Mechanisms of bubble-particle interaction are related to three different phenomena, namely, oscillating collision, liquid film creation and sliding. When the flow-driven particle is in contact with the bubble, it oscillates because of the force balance between particle inertia and surface tension, i.e., oscillating collision. Then, a liquid film is created while the particle remains on the bubble surface, i.e., liquid film creation. During existence of the liquid film, the particle slides on the bubble surface, i.e., sliding. Furthermore, the particle is attached if the liquid film is ruptured. These attachment processes are characterized by three time intervals, namely, collision, sliding, and induction time intervals.

	The collision time interval is characterized by particle oscillation on the bubble surface. This oscillating motion is described as the harmonic oscillation~\cite{Evans1954}. In the same sense, the sliding time is characterized by the interval in which the particle slides over the bubble surface. This phenomenon was modeled by assuming inertialess particle and potential flow fields~\cite{Sutherland1948,Nguyen1998} and adapting the experimental results~\cite{Dobby1986,Dobby1987}. Lastly, the induction time indicates the lasting duration of lasting the liquid film. According to Nguyen et al.~\cite{Nguyen1998}, these time intervals are interpreted as a relative position between a particle and a bubble. They adapted a concept of a critical angle to describe the induction and sliding time, which will be explained in detail in Section~\ref{sec:physics}. This study uses characteristic time and geometrical relations.

	Recently, bubble--particle interaction is solved numerically. Bubbles and nearby flow fields are solved in the volume-of-fluid (VOF) framework, whereas particles are solved in a Lagrangian framework~\cite{Li1999, Chen2004, Xu2016, Pozzetti2018, Zhao2020}. In most works, bubble--particle interactions, such as collision, sliding, and attachment, are not considered. However, such phenomena are critical factors in determining particle collecting efficiency or probability. Some {work consider} actual physical phenomena. They modeled hydrophobic forces that act on particles near a stationary bubble and concluded that considering the acting forces on particles and bubbles is critical in predicting bubble--particle interaction~\cite{Moreno2013, Moreno2016}. However, known flow fields are used, and the dynamics of a phase interface are ignored.

	In the present study, we target to model complex bubble--particle interaction by considering particle attachment mechanisms. Flow fields and bubble interfaces are solved in an Eulerian framework using the VOF method. Particle motions are solved in a Lagrangian framework. To model bubble--particle interaction, we adapt the theoretical model developed by Nguyen et al.~\cite{Nguyen1998}, which reflects the actual physical phenomena. In this model, the position of particles relative to the bubble surface is a key factor. However, in the VOF framework, a bubble interface is represented in an Eulerian framework. Thus, the exact collision point cannot be determined. To address this problem, an effective bubble is adopted to model particle motions on the bubble surface accurately, which will be demonstrated in Section~\ref{sec:numerical_modeling}. 

	The present paper begins with a description of the numerical methods for simulating gas--liquid--solid flows in Section~\ref{sec:numerical_methods}. The physical background and numerical algorithms of bubble--particle interaction are presented in Sections~\ref{sec:physics} and~\ref{sec:numerical_modeling}, respectably. Section~\ref{sec:results} provides the validation results of the present numerical method and compared with the experimental results by Hewitt et al.~\cite{Hewitt1995}. The conclusions are drawn in Section~\ref{sec:conclusions}.


	\section{Numerical methods}\label{sec:numerical_methods}	
	\subsection{Governing equations}\label{sec:NSequations}
		Gas--liquid flows are governed by the Navier--Stokes equations with surface tension. The one-fluid formulation is adopted using total density and dynamic viscosity instead of the solving governing equations of each phase. 
	The incompressible variable-density Navier--Stokes equations with surface tension are given as follows:
		\begin{equation}\label{eq:NScontinuity}
		\frac{\partial u_i}{\partial x_i}=0,
		\end{equation}
		\begin{equation}\label{eq:NSmomentum}
		\frac{\partial \left(\rho u_{i} \right)}{\partial t} + \frac{\partial \left(\rho u_{i}u_{j} \right)}{\partial x_{j}} = 
		-\frac{\partial p}{\partial x_{i}} 
		+ \frac{\partial}{\partial x_{j}}\left(\mu \left(\frac{\partial u_{i}}{\partial x_{j}} 
		+ \frac{\partial u_{j}}{\partial x_{i}} \right) \right) 
		+ \rho g_{i} + T_{i},
		\end{equation}
where $x_{i}$, $u_{i}$, $p$, $\rho$, $\mu$, $g_{i}$, and $T_{i}$ are the Cartesian coordinates, the corresponding velocity components, pressure, total density, total dynamic viscosity, gravitational acceleration, and surface tension, respectively. The total density and dynamic viscosity are defined as follows:
		\begin{equation}\label{eq:total_rho}
		\rho(c) = c\rho_g + (1 - c)\rho_l,
		\end{equation}
		\begin{equation}\label{eq:total_mu}
		\mu(c) = c\mu_g + (1 - c)\mu_l,
		\end{equation}
where $\rho_g$, $\rho_l$, $\mu_g$, $\mu_l$, and $c$ are the densities, dynamic viscosities of the gas and liquid phases, and the volume fraction of the gas phase.

	The numerical schemes for solving Eq.~(\ref{eq:NSmomentum}) are similar to those of Choi and Moin~\cite{Choi1994}. The Crank--Nicolson time advancement scheme is employed for Eq.~(\ref{eq:NSmomentum}). A Newton-iterative method is applied to solve the nonlinear terms. A second-order central difference scheme is used to discretize spatial derivative terms. Then, a fractional step method and an approximate factorization technique~\cite{Kim1985} are applied to the discretized equations. Eq.~(\ref{eq:NScontinuity}) leads to the Poisson equation for pressure, which is solved by an algebraic multigrid method~\cite{Henson2002}.

	The solid particle motions are predicted in the Lagrangian framework. The Lagrangian equations {that govern} the particle motions are expressed as follows:
		\begin{equation}\label{eq:LPT_position}
		\frac{d\mathbf{x_{p}}}{dt}=\mathbf{u_p},
		\end{equation}
		\begin{equation}\label{eq:LPT_vel}
		\frac{d\mathbf{u_p}}{dt}=\frac{f}{St_\textup{p}}(\mathbf{u-u_p}),
		\end{equation}
where $\mathbf{x_p}$, $\mathbf{u_p}$, $\mathbf{u}$, $f$, and $St_\textup{p}$ are the particle position and velocity, fluid velocity, drag coefficient, and particle Stokes number. Schiller and Naumann~\cite{Crowe2011} suggested that the drag coefficient to be modeled as follows: 
		\begin{equation}\label{eq:LPT_drag}
		f=(1+0.15{Re_\textup{p}}^{0.687}),
		\end{equation}
where $Re_\textup{p}$ is the particle Reynolds number. The particle Stokes number $St_\textup{p}$ is defined as follows:
		\begin{equation}\label{eq:LPT_St}
		St_\textup{p} = \frac{1}{18}\frac{\rho_\textup{p} {d_\textup{p}}^{2}}{\mu},
		\end{equation}
where $\rho_\textup{p}$ and $d_\textup{p}$ are the particle density and the particle diameter, respectably. Eqs~(\ref{eq:LPT_position}) and (\ref{eq:LPT_vel}) are solved using a third-order Runge--Kutta method. The fluid velocity $\mathbf{u}$ at the particle position $\mathbf{x_p}$ is obtained by employing a fourth-order Lagrange interpolating polynomial. 
	
	\subsection{Volume-of-fluid method}\label{sec:vof_method}
	
	The phases are distinguished by volume fraction and it is advected by the VOF method. The advection equations for the volume fraction $c$ is written as follows:
		\begin{equation}\label{eq:vof_advection}
		{{\partial c} \over {\partial t}} + u_i \frac{\partial c}{\partial x_i}  = 0.
		\end{equation}
A piecewise linear interface calculation (PLIC) approach~\cite{Scardovelli1999} is used  to solve Eq.~(\ref{eq:vof_advection}). In this approach, the phase interface of each cell is represented as a line or a plane equation. The VOF/PLIC method proceeds in two steps. First, the interface is reconstructed by the volume fraction $c$. The reconstruction step requires the estimation of the normal vector and the plane constant of the interface. The interface of each cell is represented as follows:
		\begin{equation}\label{eq:vof_plane}
		n_i \cdot x_i = \alpha,
		\end{equation}
where $n_i$, $x_i$, and $\alpha$ are the local normal vector of the interface, the position vector, and the plane constant of the interface{, respectably}. In the present study, the normal vector $n_i$ is calculated by a mixed Youngs-centered (MYC) method, as suggested by Aulisa et al.~\cite{Aulisa2007}. Then, the plane constant $\alpha$ can be obtained analytically using the geometrical relation between the normal vector $n_i$ and the volume fraction $c$~\cite{Scardovelli2000}. After the reconstruction step, the volume fraction $c$ is advected by Eq.~(\ref{eq:vof_advection}) using a geometrical flux computation. The conservative advection scheme suggested by Weymouth and Yue~\cite{Weymouth2010} is adopted to guarantee that the error of mass conservation is within the range of machine accuracy. The total density $\rho$ and dynamic viscosity $\mu$ are calculated with Eqs.~(\ref{eq:total_rho}) and ~(\ref{eq:total_mu}) using the updated volume fraction, respectably.

	Furthermore, surface tension $T_i$ needs to be estimated at the phase interface. This force is calculated by the continuum surface force  method of Brackbill~\cite{Brackbill1992} :
		\begin{equation}\label{eq:tension}
		T_{i}=\sigma \kappa \frac{\partial c}{\partial x_i},
		\end{equation}
where $\sigma$ and $\kappa$ are the surface tension coefficient and the curvature of the phase interface , respectably. The curvature $\kappa$ is estimated by the height function method combined with the paraboloid-fitted method developed by Popinet~\cite{Popinet2009}. The sharpness of the surface tension is conserved by implementing a balanced force algorithm proposed by Francois et al.~\cite{Francois2006}. 



	\section{Physics of bubble--particle interaction}\label{sec:physics}

	\subsection{Mechanisms of sliding and attachment of the colliding particle}\label{sec:mechanisms}
	
	The numerical modeling of bubble--particle interaction requires understanding on how a particle moves on a bubble surface. In this study, we focus on the case where the particle size is much smaller than the bubble size that is the particle diameter $d_\textup{p}$ is less than 0.1 times of the minimum bubble diameter $d_\textup{b}$, which is 0.75 mm. Moreover, a particle shape is assumed to be a rigid sphere. Under this condition, the particle momentum is negligible compared with the bubble momentum. In other words, the particle motions do not affect those of the bubble. In addition, these size differences cause capillary force to become dominant at the bubble surface. Consequently, the particle cannot transform the bubble surface. 

\begin{figure}
	\centering
	\includegraphics[width=0.9\textwidth]{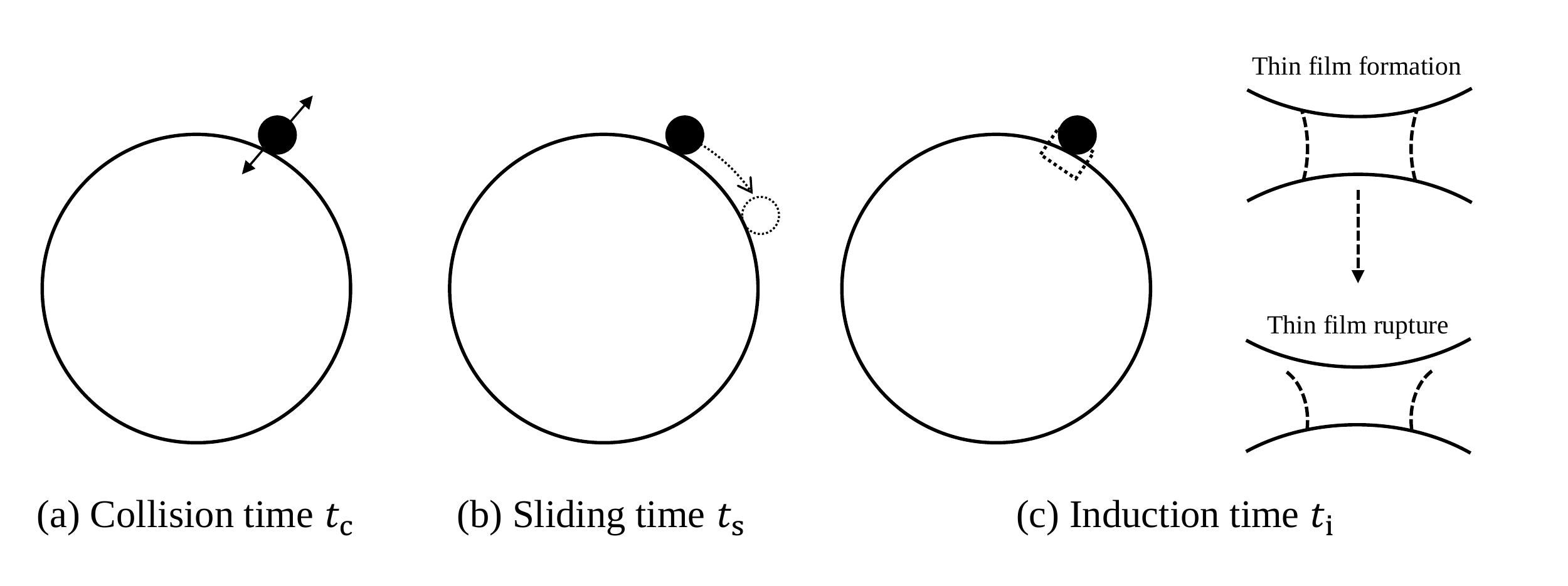}
	\caption{Schematic of three time intervals to determine particle motions on a bubble surface: (a) collision time $t_\textup{c}$; (b) sliding time $t_\textup{s}$; (c) induction time $t_\textup{i}$.}
	\label{fig:time_intervals}
\end{figure}

	The particle motions on the bubble surface can be illustrated using three characteristic time intervals. These time intervals determines whether the particle is attached or slides away after collision (Fig.~\ref{fig:time_intervals}). First, the collision time $t_\textup{c}$ refers to the duration of oscillation before sliding. Second, the sliding time $t_\textup{s}$ corresponds to the period when the particle slides over the bubble surface. Lastly, the induction time $t_\textup{i}$ covers the period from thin liquid film formation to rupture. The thin liquid film plays an important role in determining sliding or attachment of the colliding particle. {When} the thin liquid film exists, the particle keeps sliding because the thin film interrupts the attraction force between the particle and the bubble surface. In other words, the particle is attached on the condition that $t_\textup{c}$ or $t_\textup{s} > t_\textup{i}$. When $t_\textup{c} > t_\textup{i}$, the thin liquid film is ruptured during particle oscillation, and then the particle is attached. If $t_\textup{s} > t_\textup{i}$, then the thin liquid film is ruptured during sliding. When $t_\textup{s} \leqslant t_\textup{i}$, the thin liquid film exist during sliding. Finally, the particle slides away. The relationships among the time criteria are summarized as follows:

	\renewcommand{\theenumi}{\roman{enumi}}
	\begin{enumerate}
	\item If $t_\textup{c} > t_\textup{i}$, then the particle is attached during oscillation.
	\item If $t_\textup{s} > t_\textup{i}$, then attachment occurs during sliding.
	\item If $t_\textup{s} \leqslant t_\textup{i}$, then the particle slides away.
	\end{enumerate}

\begin{figure}
	\centering
	\includegraphics[width=0.9\textwidth]{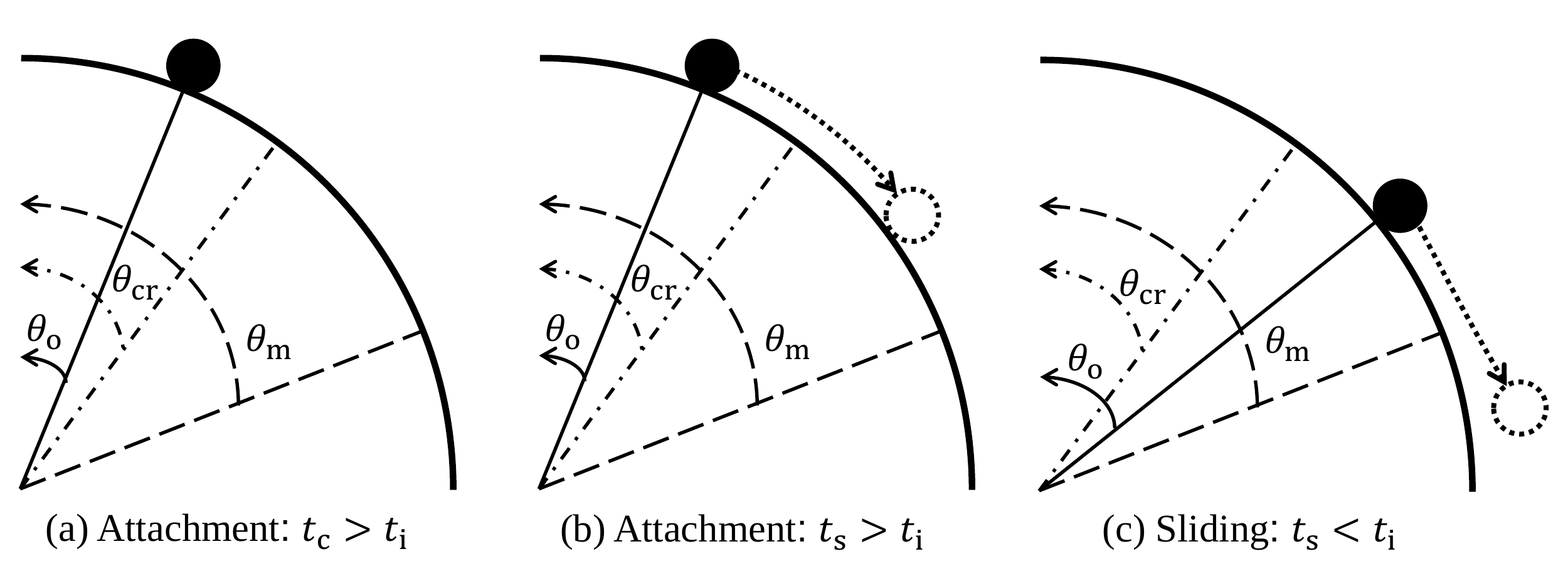}
	\caption{Schematic of three types of particle motions after collision. (a) attachment during oscillation; (b) attachment during sliding; (c) sliding away due to the thin liquid film.}
	\label{fig:angle}
\end{figure}

	Particle motions on the bubble surface can also be described {according to} relative position of the bubble and the particle. Relative position is represented by the angle measured from a stagnation point to the point of particle contact. Three possible situations after particle collision are shown in Fig.~\ref{fig:angle} using the angle criteria. Three characteristic angles are used to determine particle motions. First, the bubble--particle angle $\theta_\textup{o}$ is the angle at the particle position from the stagnation point. Secondly, the critical angle $\theta_\textup{cr}$ is the angle beyond which no attachment occurs. Lastly, the collision angle $\theta_\textup{m}$ is the angle beyond which no collision occurs. From the definition, no collision occurs when $\theta_\textup{o} > \theta_\textup{m}$. If $\theta_\textup{cr} \leqslant \theta_\textup{o} \leqslant \theta_\textup{m}$, then the particle collides but the particle slides away. The reason is that the collision point is beyond the critical angle $\theta_\textup{cr}$. When $\theta_\textup{o} < \theta_\textup{cr}$, the particle is attached. The relationships among the angle criteria are summarized as follows:
	\renewcommand{\theenumi}{\roman{enumi}}
	\begin{enumerate}
	\item If $\theta_\textup{o} > \theta_\textup{m}$, then no collision occurs.
	\item If $\theta_\textup{cr} \leqslant \theta_\textup{o} \leqslant \theta_\textup{m}$, then the particle slides away.
	\item If $\theta_\textup{o} < \theta_\textup{cr}$, then the particle is attached during oscillation or sliding.
	\end{enumerate}
%
%
%
%
In the present model, the time and angle criteria are used to determine the particle motion. The calculation procedures of the collision angle $\theta_\textup{m}$ and the critical angle $\theta_\textup{cr}$ are demonstrated in Section~\ref{sec:probability}.

\begin{figure}
	\centering
	\includegraphics[width=0.4\textwidth]{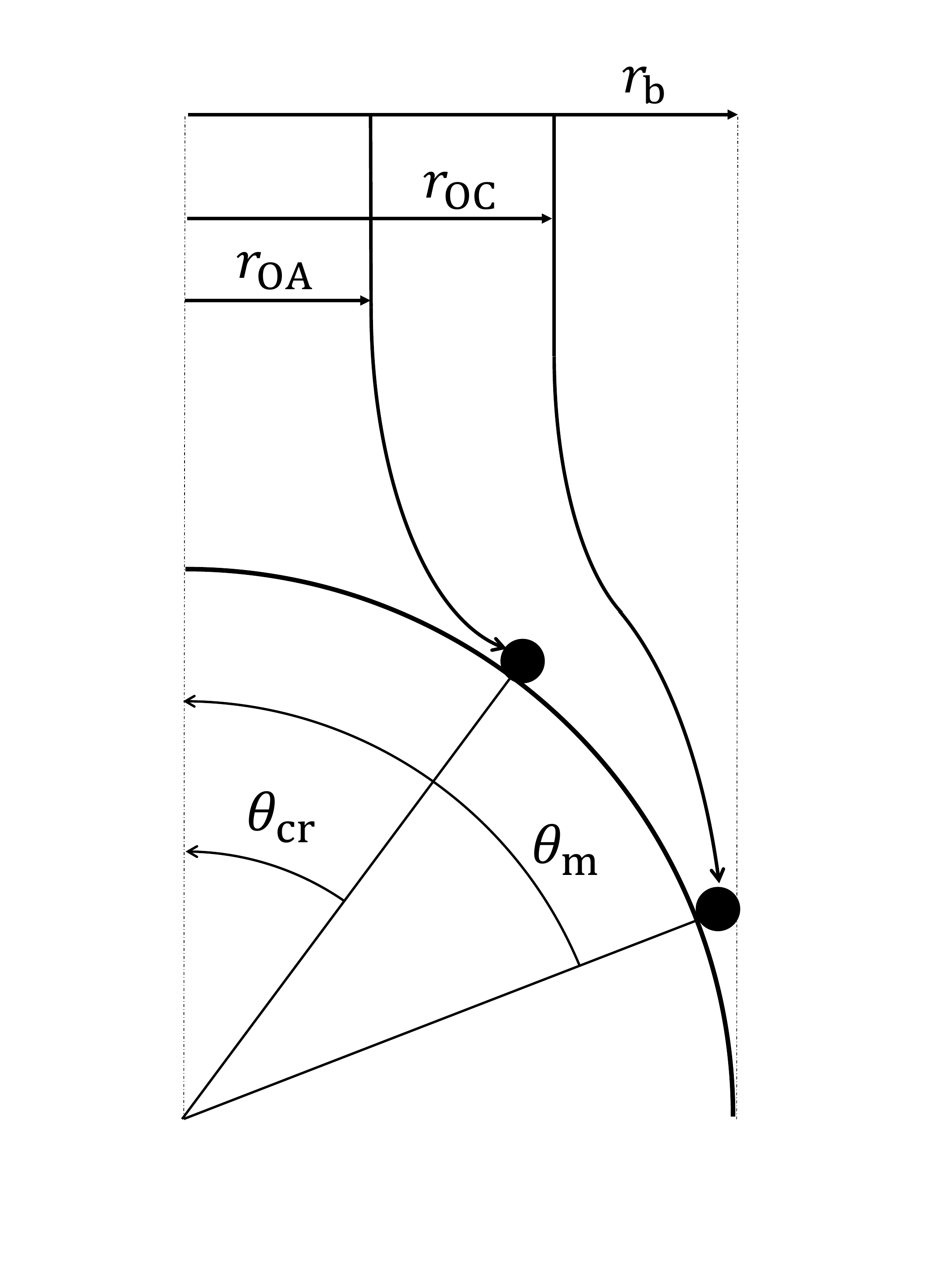}
	\caption{Schematic of the collision radius $r_\textup{OC}$ and the attachment radius $r_\textup{OA}$ with the collision angle $\theta_\textup{m}$ and the critical angle $\theta_\textup{cr}$.}
	\label{fig:radius}
\end{figure}

	\subsection{Collision and attachment probability}\label{sec:probability}
	
	According to particle motions on the bubble surface, the number of the colliding and attached particles are determined. From these numbers, a collision and attachment probability are defined. These probabilities can be used to estimate the efficiency of particle removal. Definitions and model equations are introduced in this section. In the simulation, the collision and attachment probability are directly calculated on the basis of the definition. The model equations are used as reference values. In addition, the collision angle $\theta_\textup{m}$ and the critical angle $\theta_\textup{cr}$ are calculated from the model equations.


The collision probability $P_\textup{c}$ is defined as the fraction of an actual number of colliding particles and an ideal number of colliding particles~\cite{Nguyen1998} as follows:

	\begin{equation}\label{eq:collision_prob_num}
		P_\textup{c}=\frac{N_\textup{cr}}{N_\textup{ci}},
	\end{equation}
where $N_\textup{ci}$ is the ideal number of colliding particles, and $N_\textup{cr}$ is the actual number of colliding particles. Fig.~\ref{fig:radius} schematically shows that how particles collide to the bubble surface. Ideal collision implies that all particles within the bubble sweep path collide with the bubble surface. In other words, if the location of the particles is within the bubble radius $r_\textup{b}$, then the particles ideally collide with the bubble surface. However, the actual collision occurs when the particle location is within the collision radius $r_\textup{OC}$. From this definition, the collision probability $P_\textup{c}$ can be calculated geometrically as follows:

		\begin{equation}\label{eq:collision_diameter}
		P_\textup{c}=\frac{\pi r_\textup{OC}^2}{\pi r_\textup{b}^2}=\left(\frac{r_\textup{OC}}{r_\textup{b}}\right)^2=\left(\frac{r_\textup{b}\sin{\theta_\textup{m}}}{r_\textup{b}}\right)^2,
		\end{equation}		
where ${\pi r_\textup{b}^2}$ is the area where the ideal collision occurs, and ${\pi r_\textup{OC}^2}$ is the area where the actual collision occurs.

The collision probability $P_\textup{c}$ also can be modeled using the material properties and flow conditions~\cite{Nguyen1994} as follows:
	\begin{equation}\label{eq:collision_probability}
	P_\textup{c}=\frac{2u_\textup{b} D}{9(u_\textup{b}+u_\textup{p})Y}\left(\frac{d_\textup{p}}{d_\textup{b}}\right)^2\left[\sqrt{(X+C)^2+3Y^2}+2(X+C) \right]^2,
	\end{equation}
where the dimensionless parameters $X, Y, C,$ and $D$ are given by
		\begin{subequations}
		\begin{align}
		X &=\frac{3}{2}+\frac{9Re_\textup{b}}{32+9.888{Re_\textup{b}}^{0.694}}, \label{eq:dim_x} \\
		Y &=\frac{3Re_\textup{b}}{8+1.736{Re_\textup{b}}^{0.518}}, \label{eq:dim_y} \\
		C &=\frac{u_\textup{p}}{u_\textup{b}}\left(\frac{d_\textup{b}}{d_\textup{p}} \right)^{2}, \label{eq:dim_c} \\
		D &=\frac{\sqrt{(X+C)^{2}+3Y^{2}}-(X+C)}{3Y},
		\end{align}
		\end{subequations}
where $u_\textup{b}, u_\textup{p}$, and $Re_\textup{b}=\rho_l u_\textup{b} d_\textup{b} / \mu_l$ are the bubble velocity, particle velocity, and bubble Reynolds number, respectably. To calculate Eq~(\ref{eq:collision_probability}), the bubble velocity and particle velocity need to be obtained by using material properties. The bubble velocity $u_\textup{b}$ has an empirical relation in the case of a rising single bubble~\cite{Clift1978} as follows:
	\begin{equation}\label{eq:bubble_vel}
	u_\textup{b}=0.138g^{0.82}\left(\frac{\rho_l}{\mu_l}\right)^{0.639}d_\textup{b}^{1.459} \hspace{30pt} (d_\textup{b} <1.3\textup{mm}).
	\end{equation}
The particle velocity is given by Oeters~\cite{Oeters1994} as follows:

	\begin{equation}\label{eq:particle_vel}
	u_\textup{p}=\frac{(\rho_l-\rho_\textup{p})d_\textup{p}^2 g}{18 \mu_l},
	\end{equation}
	
\noindent where $\rho_\textup{p}$ is the particle density. Eqs.~(\ref{eq:bubble_vel}) and~(\ref{eq:particle_vel}) are substituted into Eq.~(\ref{eq:collision_probability}) to obtain the collision probability. Then, the collision angle $\theta_\textup{m}$ can be calculated using Eqs.~(\ref{eq:collision_diameter}) and~(\ref{eq:collision_probability}).

	The attachment probability $P_\textup{a}$ is defined by the fraction of as actual number of attached and colliding particles~\cite{Nguyen1998} as follows: 
	
	\begin{equation}\label{eq:attachment_prob_num}
	P_\textup{a}=\frac{N_\textup{a}}{N_\textup{cr}},
	\end{equation}	
where $N_\textup{a}$ is the number of attached particles. The actual attachment occurs {when} the {particle} location is within the attachment radius $r_\textup{OA}$ {(Fig.~\ref{fig:radius})}. $P_\textup{a}$ of 1 denotes that all colliding particles are attached on the bubble surface. In the same manner of defining the collision probability, we can define the attachment probability $P_\textup{a}$ as follows:

		\begin{equation}\label{eq:attachment_diameter}
		P_\textup{a}=\frac{\pi r_\textup{OA}^2}{\pi r_\textup{OC}^2}=\left(\frac{r_\textup{OA}}{r_\textup{OC}}\right)^2=\left(\frac{r_\textup{b}\sin{\theta_\textup{cr}}}{r_\textup{b}\sin{\theta_\textup{m}}}\right)^2,
		\end{equation}		
where $\pi r_\textup{OA}^2$ is the area {where} the particles are attached. 

In this study, the attachment probability $P_\textup{a}$ is fitted from the experimental measurements conducted by Hewitt et al.~\cite{Hewitt1995}. An exponential form is adopted as follows:

		\begin{equation}\label{eq:attachment_probability}
		 P_\textup{a}=C_1\times \exp(d_\textup{p}\times C_2),
		 \end{equation}
where model constants $C_1$ and $C_2$ are written by
		\begin{subequations}\label{eq:attachment_coeff}
		\begin{align}
		C_1 &=-0.4382\times 10^{3} \times d_b + 3.9173\times \sin(\theta_\textup{cont}/2) - 0.4581, \\
		C_2 &=-31.86\times d_b -8.7973\times \sin(\theta_\textup{cont}/2) + 3.34\times 10^{-3},
		\end{align}
		\end{subequations}
where $\theta_\textup{cont}$ is the contact angle. Then, the critical angle $\theta_\textup{cr}$ can be calculated using  the collision angle $\theta_\textup{m}$, Eqs.~(\ref{eq:attachment_diameter}) and ~(\ref{eq:attachment_probability}).



	\section{Numerical modeling collision, sliding, and attachment}\label{sec:numerical_modeling}
\begin{figure}
	\centering
	\includegraphics[width=0.9\textwidth]{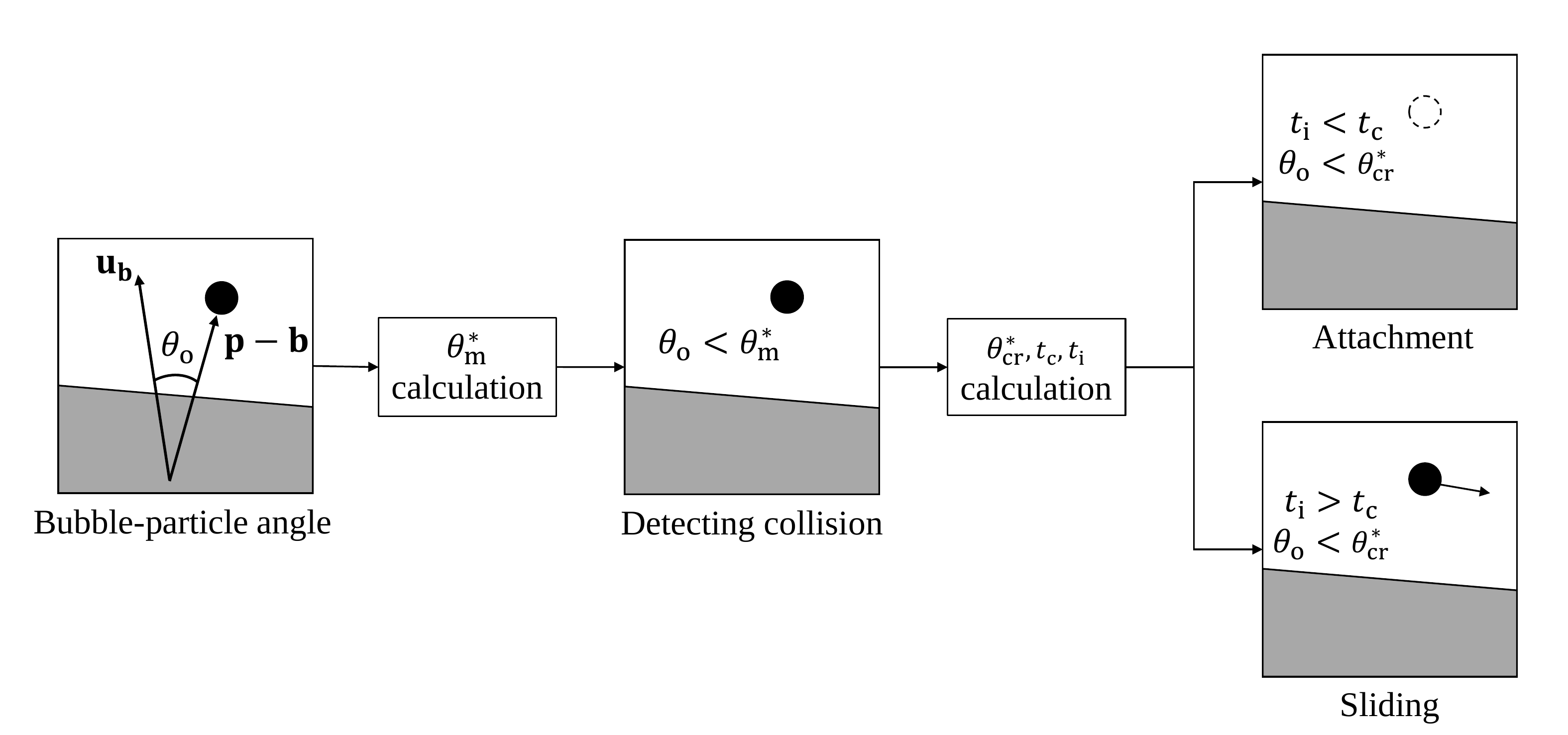}
	\caption{Schematic of total procedures of the bubble--particle interaction model. The method of calculating the bubble-particle angle $\theta_\textup{o}$ is described in Section~\ref{sec:position}. The effective contact angle $\theta_\textup{m}^{\ast}$ and critical angle $\theta_\textup{cr}^{\ast}$ are illustrated in Section~\ref{sec:effective}. The algorithm of detecting collision is presented in Section~\ref{sec:collision}. Calculation procedures of the collision time $t_\textup{c}$ and induction time $t_\textup{i}$ are found in Section~\ref{sec:numerical_determination}. The algorithm of determining sliding or attachment is explicated in Section~\ref{sec:numerical_determination}.}
	\label{fig:attachment_sliding1}
\end{figure}

	The present study mainly aims to model the physics of the bubble--particle interaction explained in Section~\ref{sec:physics} numerically. Collision occurs when the particle is in contact with the bubble surface. Then, the particle is attached or slides away according to the initial collision position of the particle. 
	Particle motions including collision, sliding, and attachment, are predicted using the time and angle criteria obtained from empirical models. The present numerical model consists of two procedures. The first one detects the collision of the particle near the bubble surface. The second procedure determines the sliding or attachment of the colliding particle. Fig.~\ref{fig:attachment_sliding1} schematically illustrates the total procedures of modeling of the bubble--particle interaction. Detailed algorithms and methodologies are presented in following subsections. 

\begin{figure}
	\centering
	\includegraphics[width=0.5\textwidth]{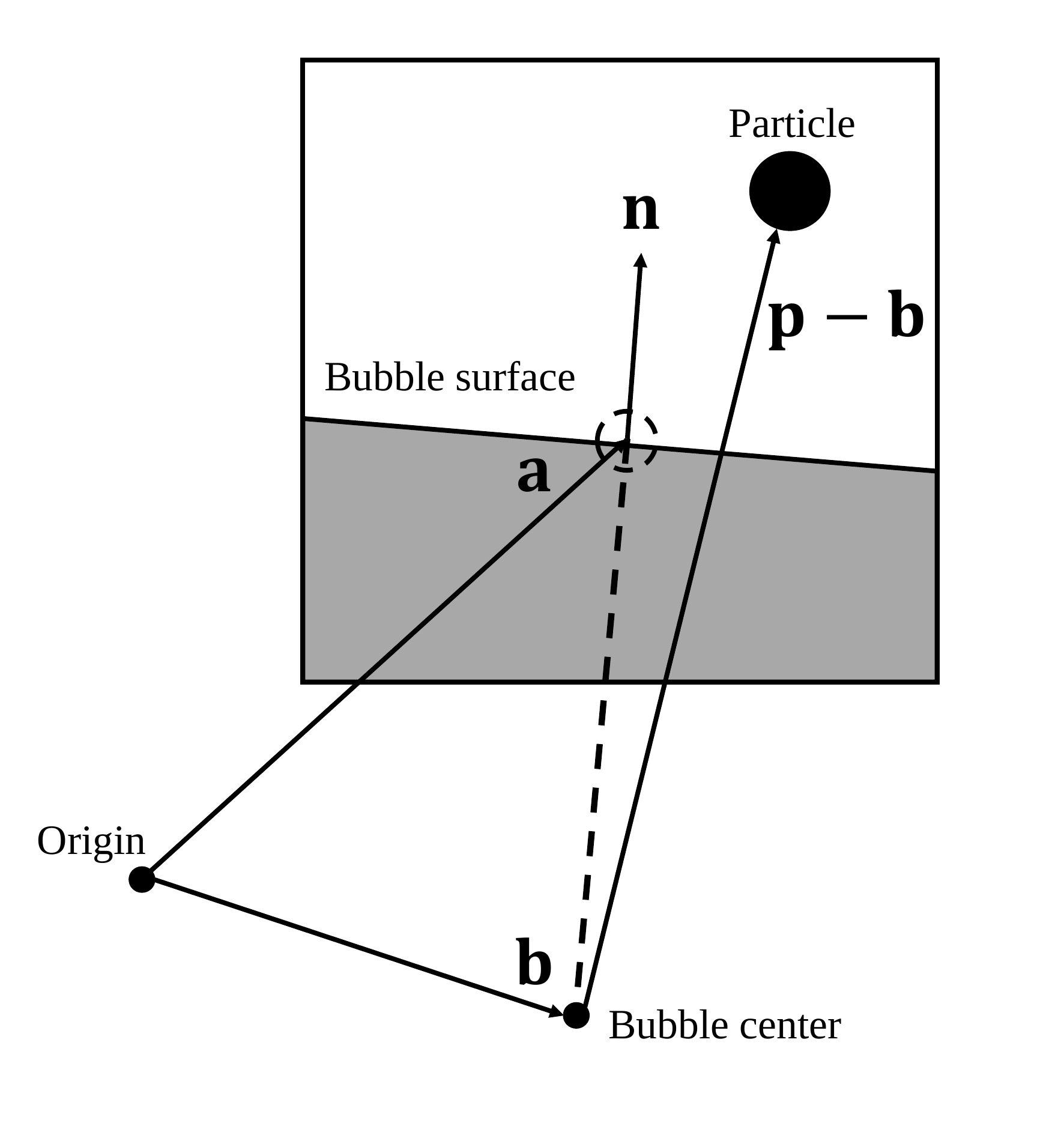}
	\caption{Schematic of position relationships between the bubble and particle. $\mathbf{a}$ is the vector from the coordinate origin to the bubble surface center. $\mathbf{n}$ is the normal vector of the bubble surface. $\mathbf{b}$ is the vector from the coordinate origin to the bubble center. $\mathbf{p}$ is the vector from the coordinate origin to the particle center. $\mathbf{p}-\mathbf{b}$ is the vector from the bubble center to the particle center.}
	\label{fig:particle_position}
\end{figure}

	\subsection{Relative position and angle between bubble and particle}\label{sec:position}
	As mentioned in Section~\ref{sec:mechanisms}, the angle criteria are {used} to determine the particle motions on the bubble surface. To apply the angle criteria, the bubble-particle angle $\theta_\textup{o}$ is required. This angle can be calculated from the position relationships of the bubble surface and the particle in a single cell (Fig~\ref{fig:particle_position}). Each vector described by the bold letter is obtained using the VOF method and simple vector calculation. The following descriptions show how to obtain the bubble-particle angle $\theta_\textup{o}$. 

	$\mathbf{n}$ is the normal vector of the bubble surface. This vector is calculated from the geometric VOF method, as described in Section~\ref{sec:vof_method}. $\mathbf{p}$ is the vector from the coordinate origin to the particle center. $\mathbf{a}$ is the vector from the coordinate origin to the center of the bubble surface, which is equal to the center of the plane equation that represents the bubble surface. This location can be found by calculating center of the plane equation. Lastly, $\mathbf{b}$ is the vector from the coordinate origin to the bubble center. Fig.~\ref{fig:particle_position} shows that the direction of the normal vector of the bubble surface $\mathbf{n}$ corresponds to $\mathbf{a}-\mathbf{b}$. The distance between the bubble surface and the bubble center is the bubble radius $r_\textup{b}$. This length is calculated from the curvature of the bubble obtained from the VOF method, that is $r_\textup{b}=2/\kappa$. By using this relation, the position vector of the bubble center $\mathbf{b}$ can be calculated as follows:
	\begin{equation}
	\mathbf{b}=\mathbf{a}-\beta \mathbf{n},
	\end{equation}
where $\beta=r_\textup{b}/|\mathbf{n}|$. 
	Finally, we can obtain the particle position vector relative to the bubble center $\mathbf{p-b}$. The bubble--particle angle $\theta_\textup{o}$ denotes the angle between the bubble velocity $u_\textup{b}$ and the particle position vector relative to the bubble center $\mathbf{p-b}$. On the basis of this definition, the bubble--particle angle $\theta_\textup{o}$ can be obtained using the inner product formula as follows:
	\begin{equation}\label{eq:bp_angle}
	\theta_\textup{o} = \arccos({\frac{\mathbf{u_b}\cdot(\mathbf{p-b})}{\left|\mathbf{u_b}\right| \left|\mathbf{p-b}\right|}}).
	\end{equation}
	
\begin{figure}
	\centering
	\includegraphics[width=0.5\textwidth]{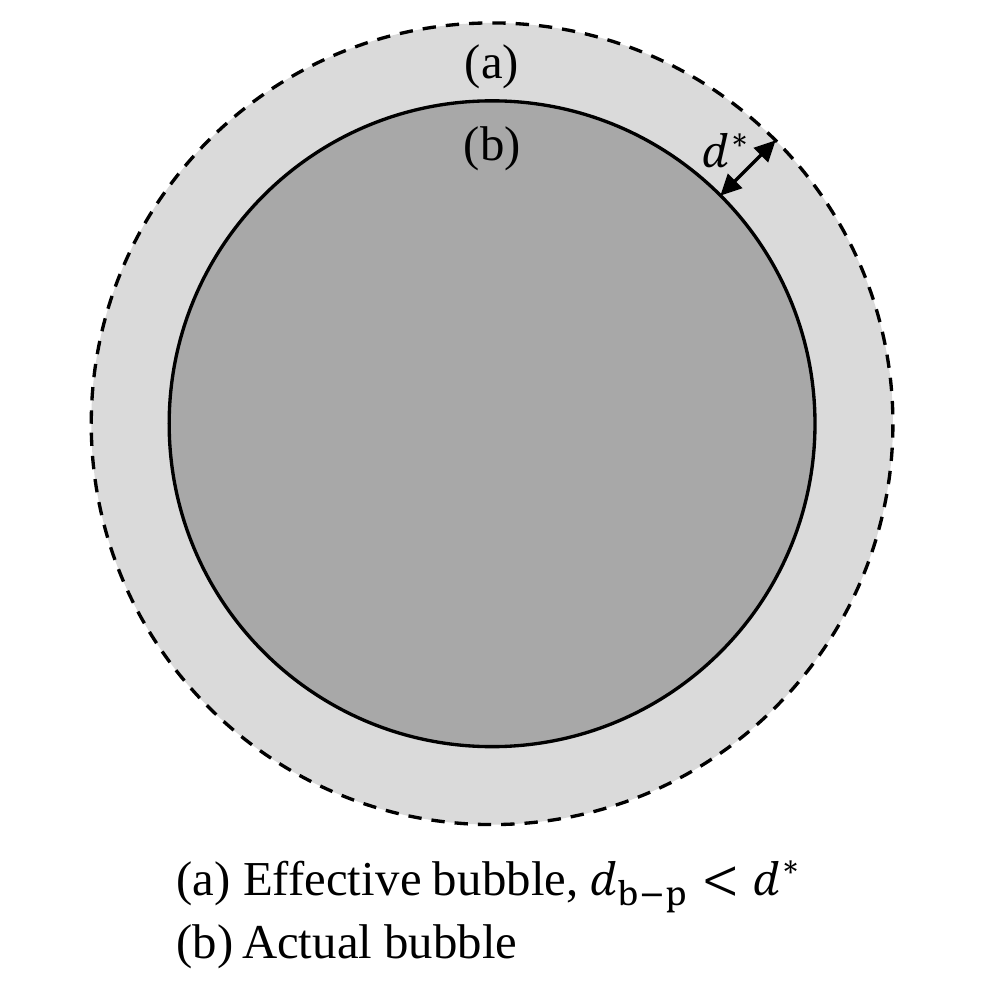}
	\caption{Schematic of (a) the effective bubble and (b) the actual bubble. $d_\textup{b-p}$ is the distance between the bubble surface to the particle center. The actual bubble has the diameter $d_\textup{b}$ and the effective bubble has the diameter $d_\textup{b}+2\times d^{\ast}$.}
	\label{fig:collision_detect2}
\end{figure}

\begin{figure}
	\centering
	\includegraphics[width=0.4\textwidth]{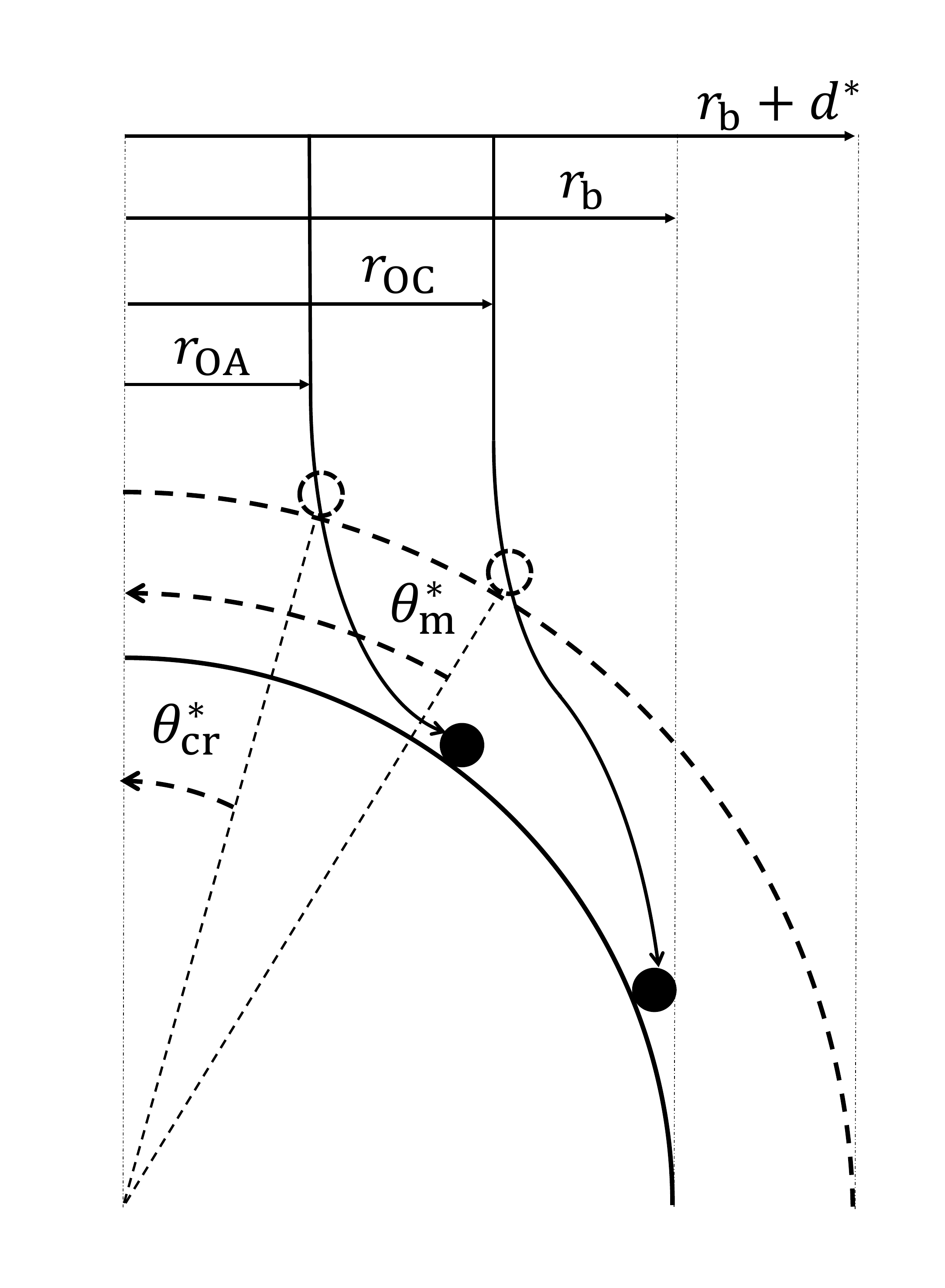}
	\caption{Schematic of the effective collision angle $\theta_\textup{m}^{\ast}$ and the effective critical angle $\theta_\textup{cr}^{\ast}$ with the collision radius $r_\textup{OC}$ and the attachment radius $r_\textup{OA}$. The solid line represents the actual bubble surface and the dashed line represents the effective bubble surface.}
	\label{fig:effective_angle}
\end{figure}

	\subsection{Effective bubble}\label{sec:detection}
	Determining particle motions on the bubble surface begins from detecting bubble--particle collision because the sliding and attachment are only examined for the colliding particle. The collision of bubble--particle implies that the particle is in a physical contact with the bubble surface. In other words, a collision condition refers to the bubble-particle distance is equal to the particle radius. However, a new definition of collision is required because the exact collision position cannot be determined for the following reason.  
	The bubble surface represented on the computational domain is not exactly defined in the VOF framework. This interface is reconstructed from the volume fraction in an Eulerian framework. However, the particle size is smaller than the grid size. Thus, the exact distance between the particle center and the bubble surface is unable to be calculated. 

	An effective bubble is introduced due to this difficulty in finding the exact position of collision. The collision is detected on the effective bubble surface. This phenomenon implies that the sliding and attachment are determined on the effective bubble surface. Fig.~\ref{fig:collision_detect2} shows the regions of two bubbles. The dark-shaded region represents the actual bubble with the diameter $d_\textup{b}$. The light-shaded region denotes the effective bubble with the diameter $d_\textup{b}+2\times d^{\ast}$. The bubble--particle distance $d_\textup{b-p}$ is defined by the distance between the particle center and the bubble surface. The bubble--particle distance $d_\textup{b-p}$ of the $N_i$-th particle can be calculated as follows:
	\begin{equation}\label{eq:distance}
	d_\textup{b-p} = \frac{\left|\mathbf{n} \cdot \mathbf{p}_{N_i} - \alpha \right|}{\left|\mathbf{n}\right|},
	\end{equation}
where $\mathbf{p}_{N_i}$ denotes the position vector of the $N_i$-th particle. The size of the effective bubble is controlled by $d^{\ast}$. 

	\subsection{Effective collision and critical angle}\label{sec:effective}
	As mentioned in the previous section, the particle motions on the bubble surface need to be determined on the effective bubble. The collision point shifts from the actual bubble to the effective bubble. Accordingly, the collision and critical angles need to be tuned. Fig.~\ref{fig:effective_angle} shows the geometrical definition of the effective collision angle $\theta_\textup{m}^{\ast}$ and the effective critical angle $\theta_\textup{cr}^{\ast}$. In this research, an effective factor $k$ is multiplied to the original collision angle $\theta_\textup{m}$ and critical angle $\theta_\textup{cr}$. The effective collision angle $\theta_\textup{m}^{\ast}$ and critical angle $\theta_\textup{cr}^{\ast}$ are defined as follows:
		\begin{subequations}
		\begin{align}
		\theta_\textup{m}^{\ast} &=k_\textup{m}\times \theta_\textup{m}, \label{eq:tunning_collision} \\
		\theta_\textup{cr}^{\ast} &=k_\textup{cr}\times \theta_\textup{cr}. \label{eq:tunning_critical}
		\end{align}
		\end{subequations}
~\ref{sec:tunning} presents the detailed form of effective factors $k_\textup{m}$ and $k_\textup{cr}$ and tuning process.

	\subsection{Detecting collision}\label{sec:collision}
	The algorithm for detecting collision is presented in the VOF framework. Algorithm~\ref{alg:collision_detection} summarizes the algorithm for detecting collision. The interface cell corresponds to the cell with the volume fraction $c$ of $0<c<1$. First, the existence of the interface is examined for each particle after updating the particle position.  A $3\times 3\times 3$ stencil is the candidate cell for examination. If the interface cell exists, then the normal vector $\mathbf{n}$ of the interface and the position of the bubble center are obtained using the geometrical VOF method. Next, the bubble--particle distance $d_\textup{b-p}$ is calculated using Eq.~(\ref{eq:distance}). The condition of $d_\textup{b-p} \leqslant d^{\ast}$ implies {that} the particle is within the effective bubble. For this case, the bubble--particle angle $\theta_\textup{o}$ and the effective collision angle $\theta_\textup{m}^{\ast}$ are calculated. The final collision is determined by comparing these two angles. If $\theta_\textup{o} \leqslant \theta_\textup{m}^{\ast}$, then the colliding particle is tagged.

	\begin{algorithm}[h]
	 {
	 	\label{alg:collision_detection}
	 	\caption{Detecting collision}
	 	\begin{algorithmic}
		 	\FOR{$i=1$ to $N_i$} 
 				\FOR{$N_i$ $cell$ and neighbor $cells$}
 					\IF{(interface $cell$)}
		 				\STATE obtain the normal vector $\mathbf{n}$ of interface and the bubble center
		 				\STATE calculate $d_\textup{b-p}$
	 					\IF{($d_\textup{b-p} \leqslant d^{\ast}$)}
	 						\STATE calculate the bubble--particle angle $\theta_\textup{o}$ and the effective collision angle $\theta_\textup{m}^{\ast}$
	 						\IF{($\theta_\textup{o} \leqslant \theta_\textup{m}^{\ast}$)}
			 					\STATE $N_i \to$ collision and tagged
		 					\ENDIF
						\ENDIF
		 			\ENDIF
		 		\ENDFOR
			\ENDFOR
	 	\end{algorithmic}
	 }
	\end{algorithm}

	\subsection{Determination of sliding or attachment}\label{sec:numerical_determination}
	After detecting collision, the last step is {to determine} sliding or attachment of the colliding particle using the time and angle criteria.  
	First, the collision time $t_\textup{c}$ and the induction time $t_\textup{i}$ are computed. The collision time $t_\textup{c}$ modeled by Evans~\cite{Evans1954} is used. This model is nearly identical to the mean value of other collision time models~\cite{Zhang2000}. The collision time $t_\textup{c}$ is modeled as follows:
		\begin{equation}\label{eq:collision_time}
		t_\textup{c}=\left(\frac{\pi^{2}\rho_\textup{p}}{12 \sigma} \right)^{\frac{1}{2}}{d_\textup{p}}^{\frac{3}{2}}.
		\end{equation}	
This time refers to the duration of oscillation after particle collision. 	
The induction time $t_\textup{i}$ modeled by Nguyen et al.~\cite{Nguyen1998} is written as follows:
		\begin{equation}\label{eq:induction_time}
		t_\textup{i}=\frac{d_\textup{p}+d_\textup{b}}{2u_\textup{b}(1-B^{2})A} \ln \left \{ \frac{\tan(\frac{\theta_\textup{m}}{2})}{\tan(\frac{\theta_\textup{cr}}{2})}  \left [\frac{\csc\theta_\textup{m}+B\cot\theta_\textup{m}}{\csc\theta_\textup{cr}+B\cot\theta_\textup{cr}}  \right ]^{B} \right \},
		\end{equation}
where dimensionless parameters A and B are given by
		\begin{subequations}
		\begin{align}
		A &=\frac{u_\textup{p}}{u_\textup{b}} + \frac{d_\textup{p}}{d_\textup{b}}X - \left(\frac{d_\textup{p}}{d_\textup{b}} \right)^{2}\left[ \frac{9}{4}+\frac{27}{64}Re_\textup{b} - 0.2266{Re_\textup{b}}^{1.1274} \right], \label{eq:dim_a} \\
		B &=\frac{d_\textup{p}}{d_\textup{b}}\frac{Y}{A} - \frac{0.437}{A}\left(\frac{d_\textup{p}}{d_\textup{b}} \right)^{2}Re_\textup{b}^{1.0562}, \label{eq:dim_b} 
		\end{align}
		\end{subequations}
where $X$ and $Y$ are given by Eqs.~(\ref{eq:dim_x}) and~(\ref{eq:dim_y}){, respectably}. The effective critical angle $\theta_\textup{cr}^{\ast}$ is obtained by Eq.~(\ref{eq:tunning_critical}).

	\begin{algorithm}[h]
	 {
	 	\label{alg:attachment_sliding}
	 	\caption{Determining sliding or attachment}
	 	\begin{algorithmic}
			\IF{($N_i$ is determined as collision)} 
			\STATE calculate the effective critical angle $\theta_\textup{cr}^{\ast}$, the collision time $t_\textup{c}$, and the induction time $t_\textup{i}$
			 	\IF{($t_\textup{i} < t_\textup{c}$)}
			 		\STATE $N_i \to$ attachment, elimination
				\ELSE
					\IF{($\theta_\textup{o} < \theta_\textup{cr}^{\ast}$)} 
					     \STATE $N_i \to$ attachment, elimination
				     \ELSE
					     	\STATE $N_i \to$ sliding
				     \ENDIF
			     	\ENDIF				
			\ENDIF
	 	\end{algorithmic}
	 }
	\end{algorithm}

	The procedures for determining sliding or attachment is summarized in Algorithm~\ref{alg:attachment_sliding}. As discussed in Section~\ref{sec:mechanisms}, $t_\textup{i} < t_\textup{c}$ indicates {that} the colliding particle is attached to the bubble surface because the thin liquid film has been ruptured during oscillation. Otherwise, the bubble--particle angle $\theta_\textup{o}$ and the effective critical angle $\theta_\textup{cr}^{\ast}$ are compared. When $\theta_\textup{o} < \theta_\textup{cr}^{\ast}$, the particle is turned out attachment and eliminated in the computational domain. 


	\section{Results}\label{sec:results}

\begin{figure}
	\centering
	\includegraphics[width=0.4\textwidth]{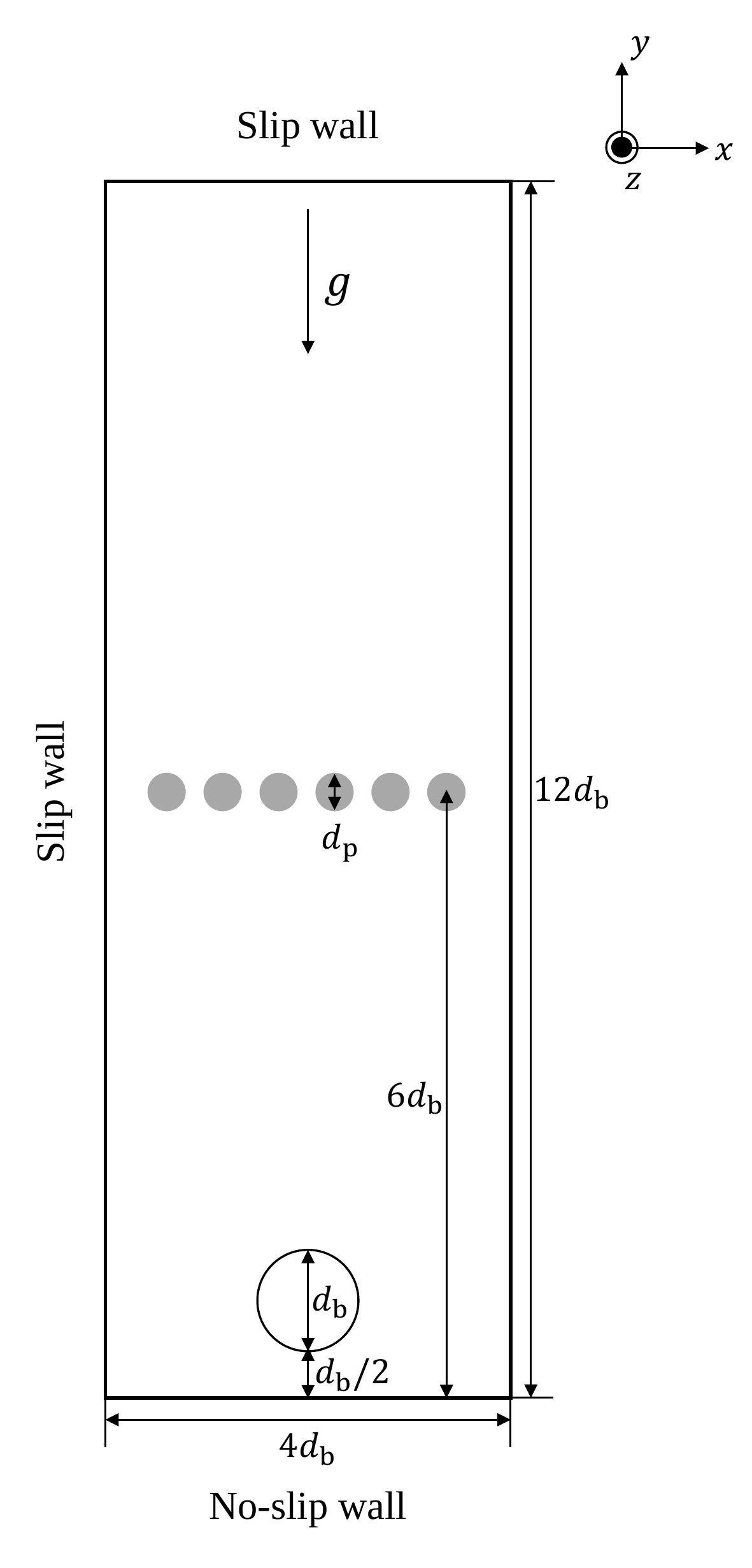}
	\caption{Computational configuration for numerical simulations of the rising of a single bubble with particles.}
	\label{fig:config1}
\end{figure}

	\subsection{Rising of a single bubble with particles}\label{sec:single}
	
\begin{figure}
	\centering
	\includegraphics[width=90mm]{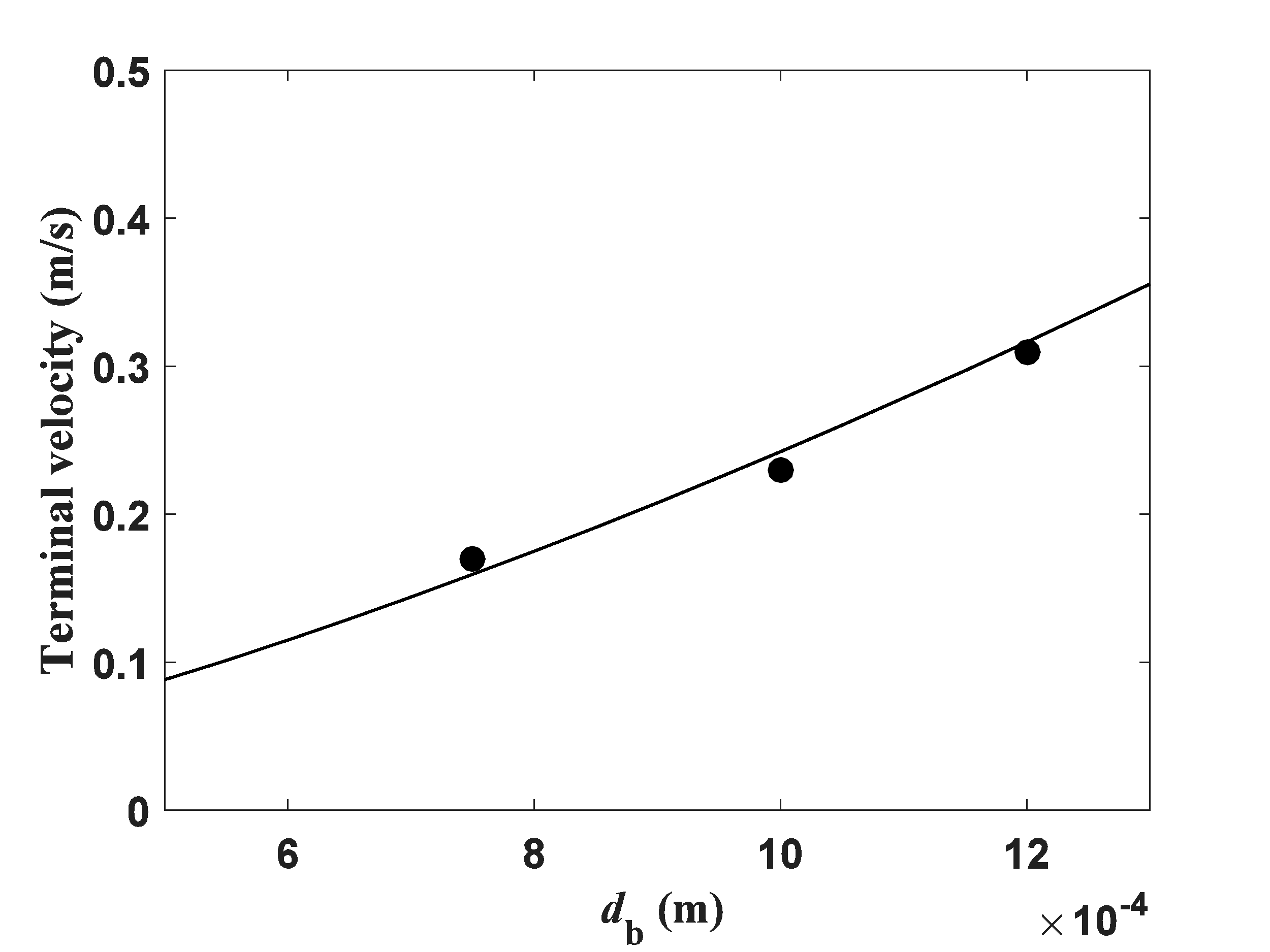}
	\caption{Terminal velocities of the rising of a single bubble as a function of the bubble diameter $d_\textup{b}$. --- : empirical results obtained from Eq.~(\ref{eq:bubble_vel}); $\bullet$ : simulation results of $d_\textup{b} =$ 0.75, 1.0, and 1.2 mm.}
	\label{fig:bubble_terminal}
\end{figure}

{\renewcommand{\arraystretch}{1.5}
\begin{table}[h]
\centering
\begin{tabular}{llllllllllll}
\hline
\multirow{2}{*}{$d_\textup{b}$ (mm) } & \multicolumn{5}{l}{$\theta_\textup{cont}=50^{\circ}$ } &  & \multicolumn{5}{l}{$\theta_\textup{cont}=88^{\circ}$} \\ \cline{2-6} \cline{8-12} 
              								     & \multicolumn{5}{l}{$d_\textup{p}$ ($\upmu$m)}           &  & \multicolumn{5}{l}{$d_\textup{p}$ ($\upmu$m)}    \\ \hline
0.75            						           & 35   & 40    & 50    & 60   & 70  					     &   & 35   & 40   & 50   & 60   & 70     \\
1.2            								& 35   & 40    & 50    & 60    & 70     						&   & 35   & 40   & 50   & 60   & 70     \\ \hline
\end{tabular} 
\caption{Simulation conditions.}
\label{tab:conditions}
\end{table}}

	Numerical simulations of the rising of a single bubble with particles are conducted to verify the present gas--liquid--solid flow solver using the bubble--particle interaction model. Fig.~\ref{fig:config1} shows the computational configuration. The simulation results are compared with the experimental results obtained by Hewitt et al.~\cite{Hewitt1995}. 
	The computational domain is defined by $0 \leqslant x,z \leqslant 4d_\textup{b}$ and $0 \leqslant y \leqslant 12d_\textup{b}$. A no-slip wall boundary condition is applied to the bottom wall. Slip wall boundary conditions are imposed to the other walls. An uniformly-distributed $64 \times 192 \times 64$ grid is utilized.
	The material properties follow the experimental conditions. The densities and dynamic viscosities of pure water and air at room temperature are considered. The surface tension coefficient is 0.074 N/m. The particle density is 2650 kg/$\textup{m}^{3}$. Parameters follow the experimental cases. Table~\ref{tab:conditions} shows the simulation conditions. $d_\textup{p}=40$ and 60 $\upmu$m cases are also considered in the simulation. 
	A stationary spherical bubble is initialized with the center positioned $d_\textup{b}$ away from the bottom wall. The initialization process is based on the numerical integration of an implicit function developed by Bn$\acute{\textup{a}}$ et al.~\cite{Bna2015}. A total of 401 particles are uniformly distributed along the $x$-direction in the range of $d_\textup{b} \leqslant x \leqslant 3d_\textup{b}$ and at the middle of the $yz$-plane. The terminal velocity of a bubble is reached before passing through particles. 
	
	 Before validating the bubble--particle interaction model, the terminal velocity of a rising single bubble without a particle is compared with that in the empirical relation given by Eq.~(\ref{eq:bubble_vel}). Terminal velocity is estimated by the rising velocity of the bubble center. The simulation results are in good agreement with the empirical relation (Fig.~\ref{fig:bubble_terminal}). The following subsections show results of the rising of the single bubble with particles.

	\subsubsection{Collision probability and diameter}\label{sec:collision_results}

\begin{figure}
	\centering
	\includegraphics[width=90mm]{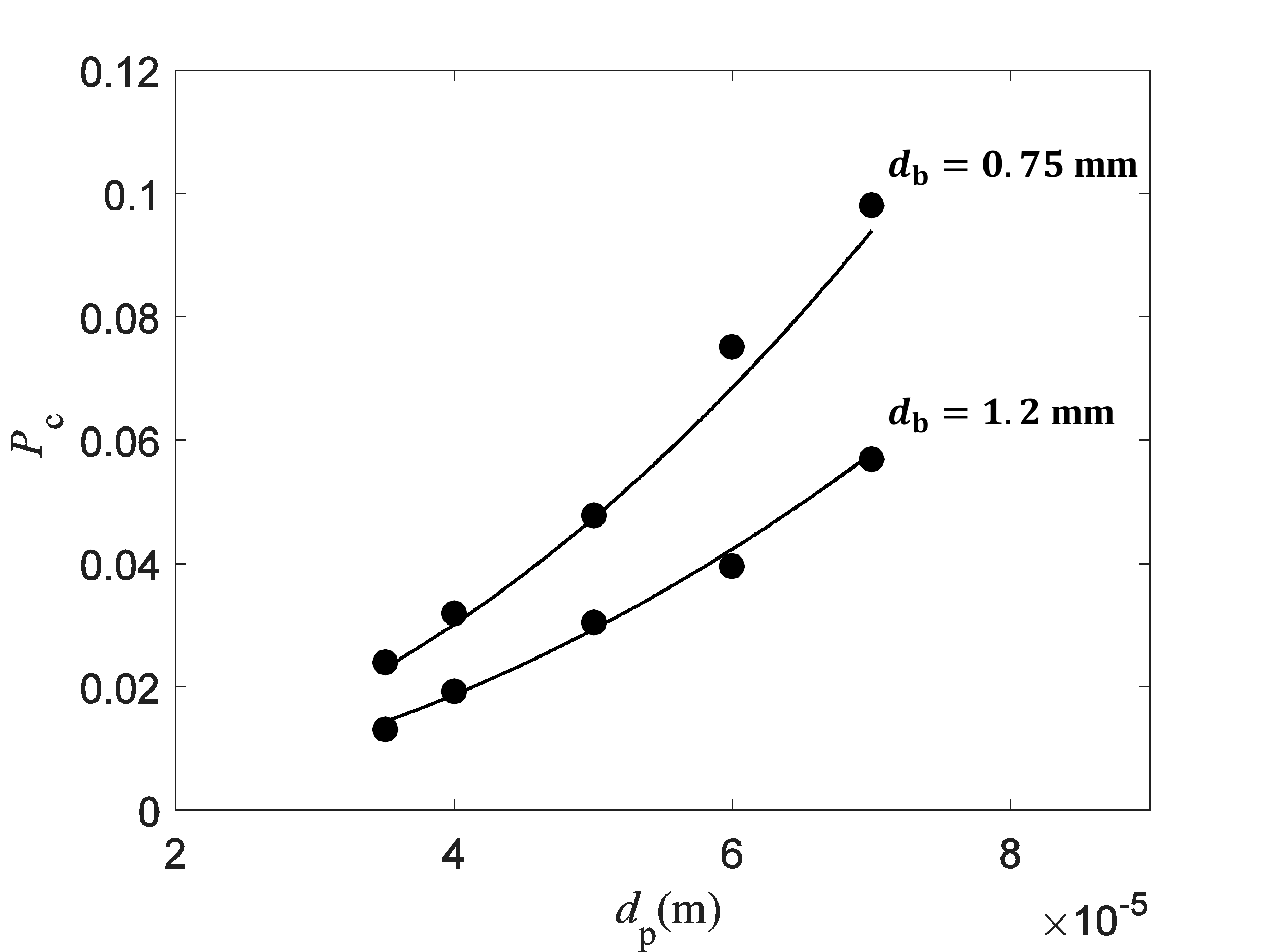}
	\caption{Collision probability $P_\textup{c}$ as a function of the particle diameter $d_\textup{p}$ for the bubble diameter $d_\textup{b}=0.75$ and 1.2 mm. --- : model depicted by Eq.~(\ref{eq:collision_probability}); $\bullet$ : simulation results of the particle diameter $d_\textup{p}=35, 40, 50, 60,$ and 70 $\upmu$m.}
	\label{fig:collision_prob}
\end{figure}
	
	The collision probabilities obtained from the present simulation using Eq.~(\ref{eq:collision_prob_num}) are compared with the model written in Eq.~(\ref{eq:collision_probability}). 
The ideal number of colliding particles $N_\textup{ci}$ is $201^2$. The particles are squared because they are linearly distributed. Similarly, the actual number of colliding particles $N_\textup{cr}$ is obtained by computing the square of the total sum of the tagged particles. Fig.~\ref{fig:collision_prob} shows the collision probabilities as a function of particle and bubble diameters. The simulation results are in good agreement with the model results. This finding shows that the present model can predict the collision probability using the effective bubble. 
	 

\begin{figure}
	\centering
	\includegraphics[width=90mm]{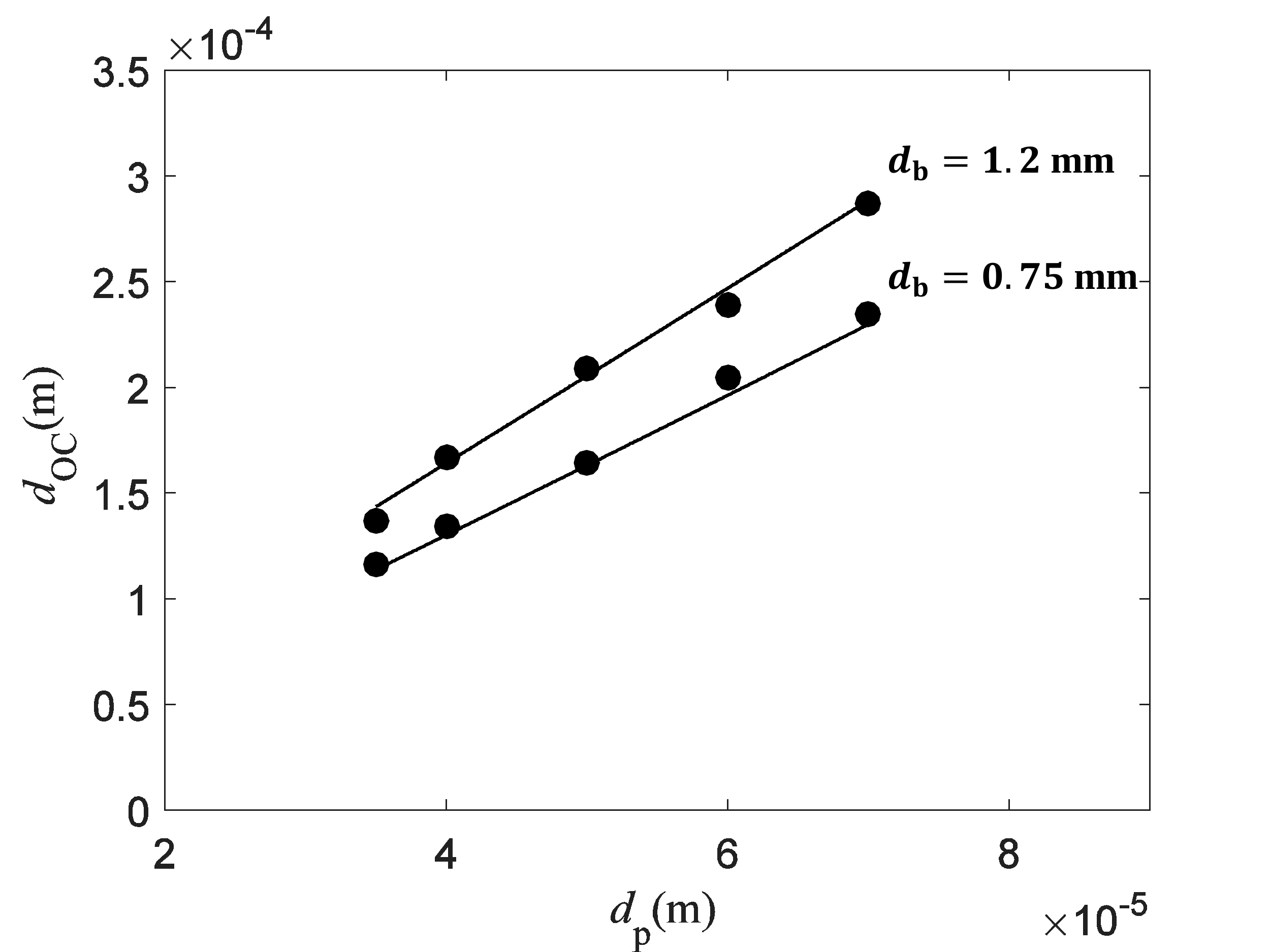}
	\caption{Collision diameter $d_\textup{OC}$ as a function of the particle diameter $d_\textup{p}$ for the bubble diameter $d_\textup{b}=0.75$ and 1.2 mm. --- : model results obtained from Eqs.~(\ref{eq:collision_diameter}) and~(\ref{eq:collision_probability}); $\bullet$ : simulation results of the particle diameter $d_\textup{p}=35, 40, 50, 60,$ and 70 $\upmu$m.}
	\label{fig:collision_dia}
\end{figure}

The collision diameters $d_\textup{OC}$ are further compared with the results calculated using the model equation. Eqs.~(\ref{eq:collision_prob_num}) and~(\ref{eq:collision_diameter}) show that the square of the collision diameter is proportional to the total number of colliding particles. This length can be obtained using Eq.~(\ref{eq:collision_diameter}), that is $d_\textup{OC}=d_\textup{b}\sqrt{P_\textup{c}}$. Fig.~\ref{fig:collision_dia} shows {that} the collision diameters as a function of particle and bubble diameters. The collision diameter $d_\textup{OC}$ linearly increases with the particle diameter $d_\textup{p}$. This trend implies that the total number of colliding particles is proportional to the square of the particle diameter $d_\textup{p}$.

	\subsubsection{Attachment probability and diameter}\label{sec:attachment_results}

\begin{figure}
	\centering
	\includegraphics[width=140mm]{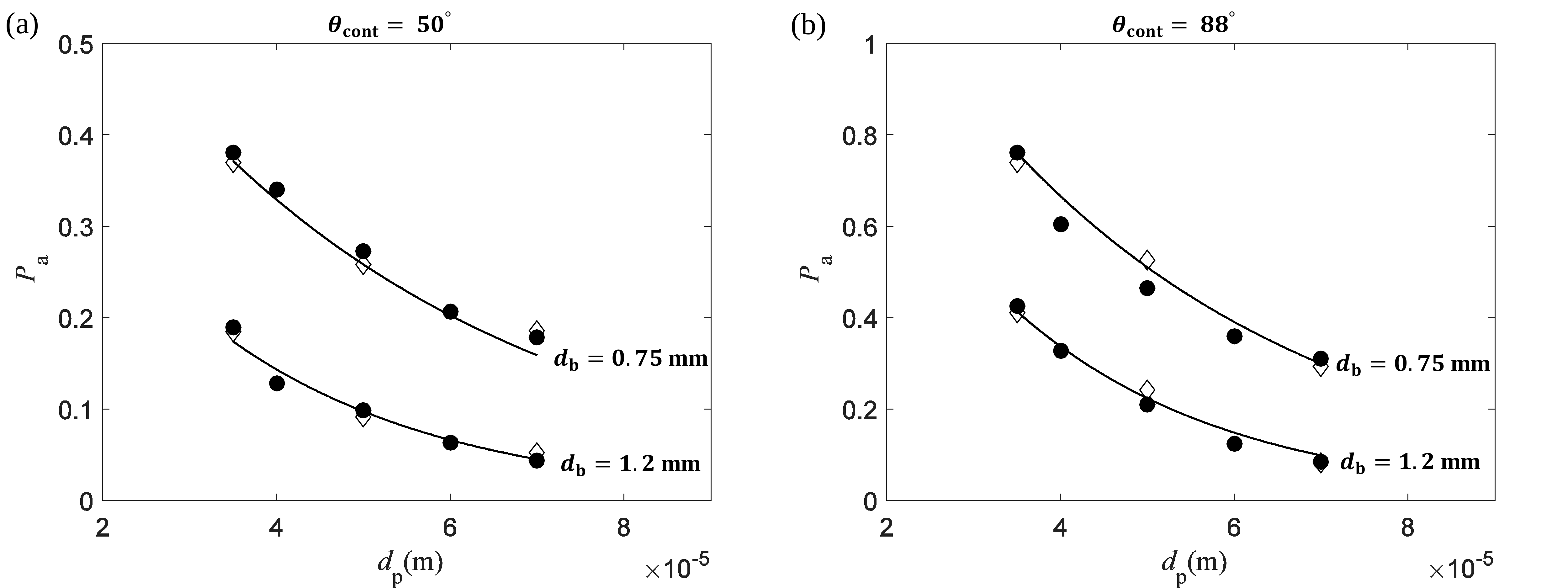}
	\caption{Attachment probability $P_\textup{a}$ as a function of the particle diameter $d_\textup{p}$ and the bubble diameter $d_\textup{b}=0.75$ and 1.2 mm for the contact angle (a) $\theta_\textup{cont}=50^{\circ}$ and (b) $88^{\circ}$. Symbols are results of the particle diameter $d_\textup{p}=35, 40, 50, 60,$ and 70 $\upmu$m. --- : fitting results obtained from Eqs.~(\ref{eq:attachment_probability}) and~(\ref{eq:attachment_coeff}); $\lozenge$ : experimental results; $\bullet$ : simulations results.}
	\label{fig:attachment_prob}
\end{figure}

The attachment probability and diameter are compared with the model and experimental results. The attachment probability in the present simulation is estimated by using Eq.~(\ref{eq:attachment_prob_num}). The total number of attached particle $N_\textup{a}$ is obtained by computing the square of the total sum of eliminated particles. Fig.~\ref{fig:attachment_prob} shows that the attachment probabilities as a function of particle diameters, bubble diameters, and contact angles. The simulation results are in favorable agreement with the experimental results. This {finding} shows {that} the effective bubble can be applied in predicting sliding and attachment. 


\begin{figure}
	\centering
	\includegraphics[width=140mm]{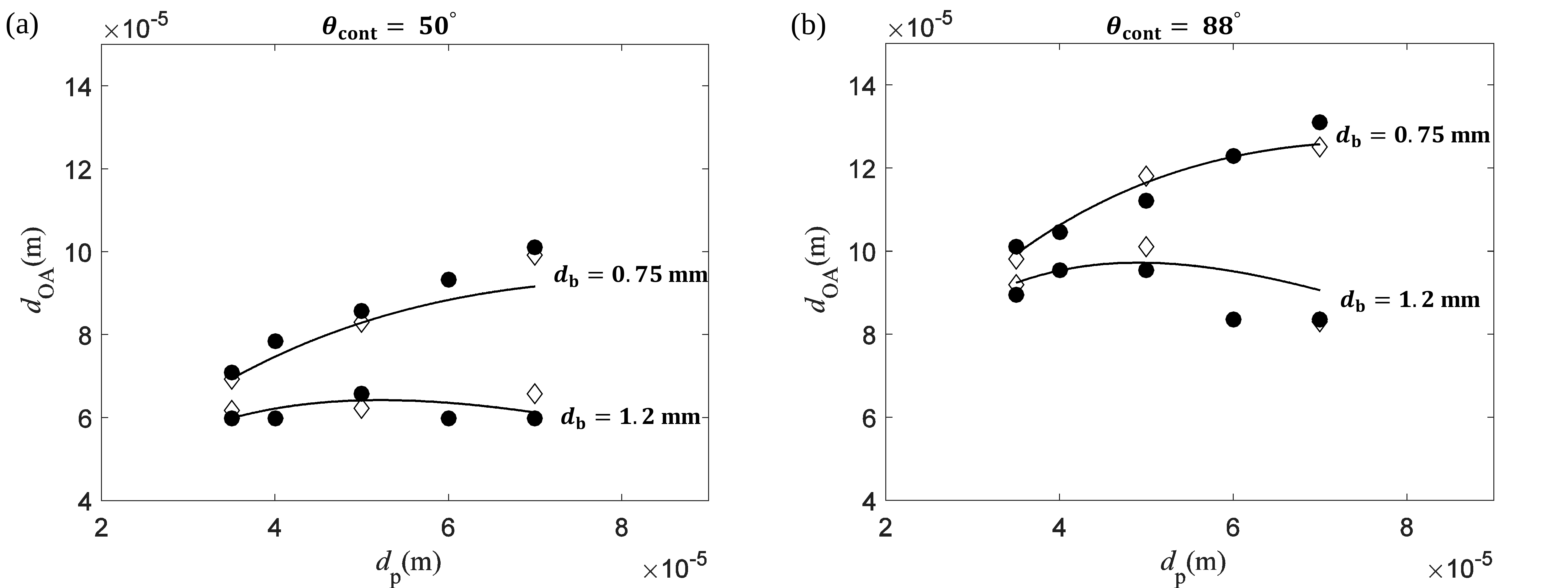}
	\caption{Attachment diameter $d_\textup{OA}$ as a function of the particle diameter $d_\textup{p}$ and the bubble diameter $d_\textup{b}=0.75$ and 1.2 mm for the contact angle (a) $\theta_\textup{cont}=50^{\circ}$ and (b) $88^{\circ}$. Symbols are results of the particle diameter $d_\textup{p}=35, 40, 50, 60,$ and 70 $\upmu$m. --- : model results; $\lozenge$ : experimental results; $\bullet$ : simulations results.}
	\label{fig:attachment_dia}
\end{figure}


The attachment diameter is calculated by $d_\textup{OA}=d_\textup{OC}\sqrt{P_\textup{a}} = d_\textup{b}\sqrt{P_\textup{c}P_\textup{a}}$ using Eqs.~(\ref{eq:collision_diameter}) and~(\ref{eq:attachment_diameter}). The square of the attachment diameter can be interpreted as the total number of attached particles. Fig.~\ref{fig:attachment_dia} shows the attachment diameters as a function of particle diameters, bubble diameters, and contact angles. Overall, the simulation results are well matched with the model and experimental results.  
When the bubble diameter is 0.75 mm, the attachment diameter $d_\textup{OA}$ increases with the particle diameter $d_\textup{p}$. 
However, for the condition of the bubble diameter of 1.2 mm, the attachment diameter does not change much as the particle diameter $d_\textup{p}$ increases.

This tendency is {due to} the product of the collision and attachment probability as follows:

	\begin{equation}\label{eq:pcpa}
		P=P_\textup{c}P_\textup{a}=\frac{N_\textup{cr}}{N_\textup{ci}}\frac{N_\textup{a}}{N_\textup{cr}}=\frac{N_\textup{a}}{N_\textup{ci}}=(\sin{\theta_\textup{cr}})^2,
	\end{equation}
where $P$ is called a flotation probability~\cite{Nguyen1998} or total attachment probability~\cite{Zhang2000}. This probability corresponds to the fraction of the number of attached particles and the ideal number of colliding particles. This relation also shows that the attachment diameter $d_\textup{OA}$ is proportional to $\sin{\theta_\textup{cr}}$ when the bubble diameter $d_\textup{b}$ is fixed. The model results of Fig.~\ref{fig:attachment_dia} is obtained from $d_\textup{b}\sqrt{P}$. Moreover, the attachment diameter $d_\textup{OA}$ increases with the contact angle $\theta_\textup{cont}$ {because} the increment of the contact angle $\theta_\textup{cont}$ causes the attachment probability $P_\textup{a}$ to rise (Fig.~\ref{fig:attachment_prob}). 

	\subsubsection{Particle trajectory}\label{sec:visualization_single}

\begin{figure}
	\centering
	\includegraphics[width=0.55\textwidth]{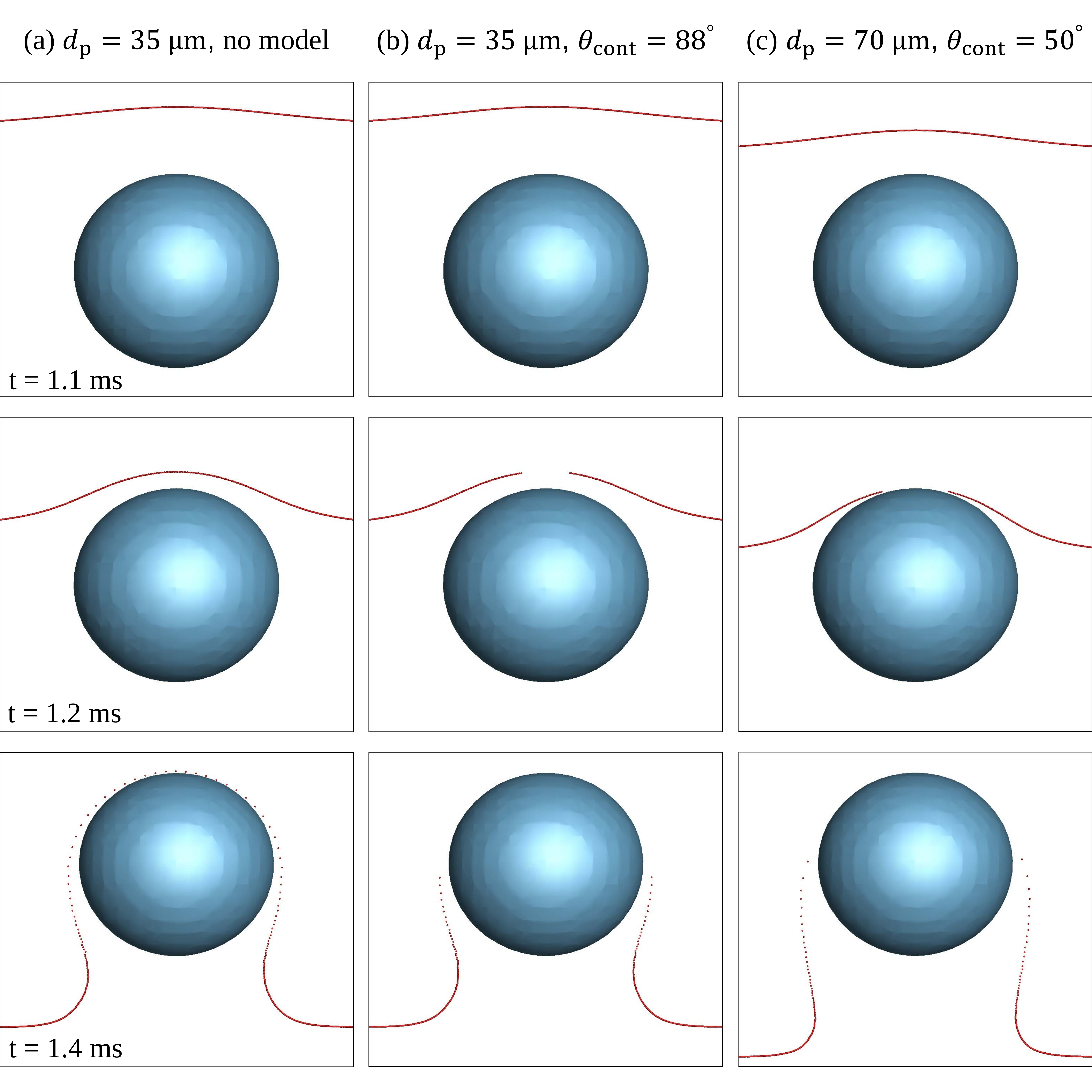}
	\caption{Particle trajectories near the bubble for $d_\textup{b}=0.75$ mm under three conditions: (a) $d_\textup{p}=35$ $\upmu$m, without the bubble--particle interaction model; (b) $d_\textup{p}=35$ $\upmu$m and $\theta_\textup{cont}=88^{\circ}$; (c) $d_\textup{p}=70$ $\upmu$m and $\theta_\textup{cont}=50^{\circ}$ at the same time and location.}
	\label{fig:visual_075}
\end{figure}
\begin{figure}
	\centering
	\includegraphics[width=0.55\textwidth]{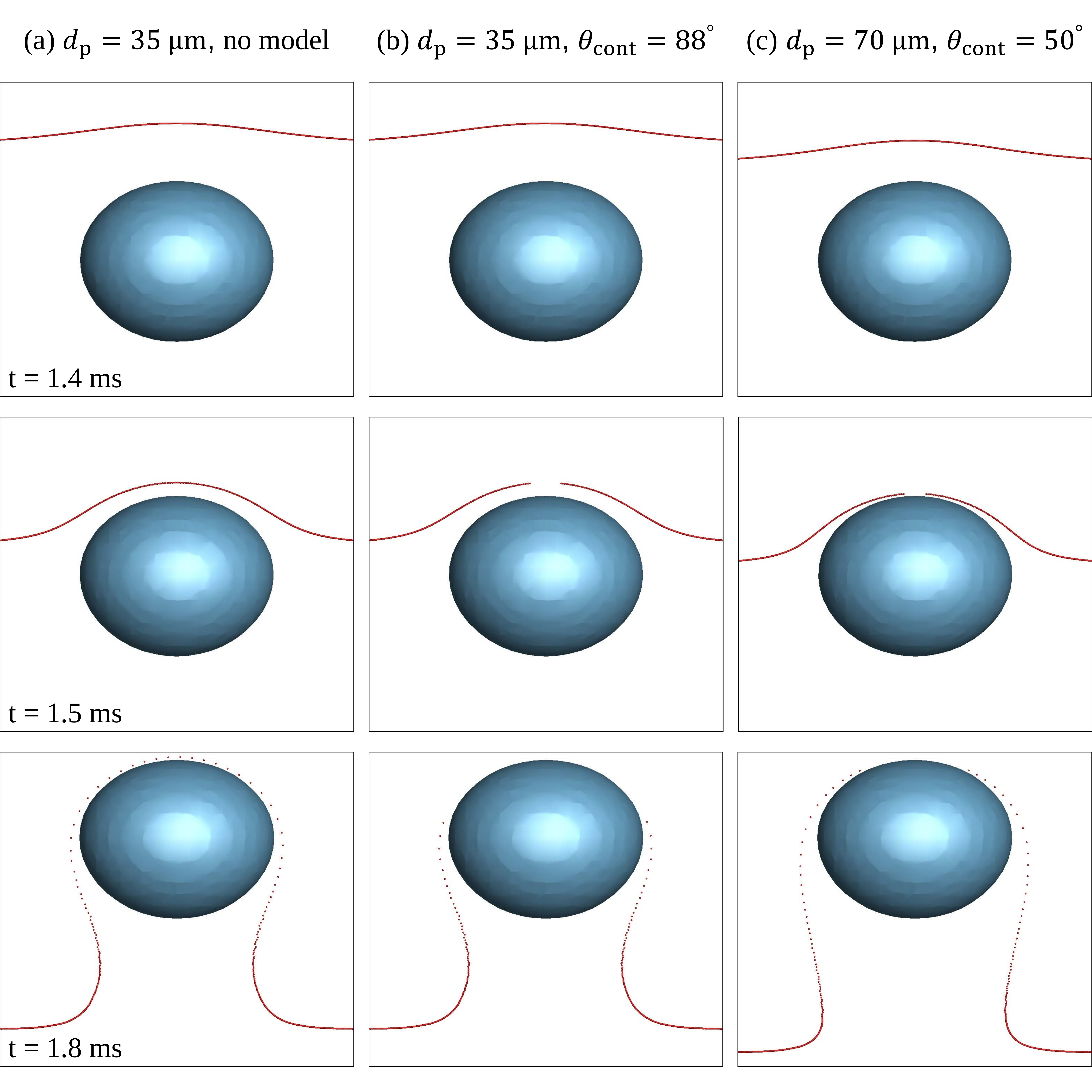}
	\caption{Particle trajectories near the bubble for $d_\textup{b}=1.2$ mm under three conditions: (a) $d_\textup{p}=35$ $\upmu$m, without the bubble--particle interaction model; (b) $d_\textup{p}=35$ $\upmu$m and $\theta_\textup{cont}=88^{\circ}$; (c) $d_\textup{p}=70$ $\upmu$m and $\theta_\textup{cont}=50^{\circ}$ at the same time and location.}
	\label{fig:visual_12}
\end{figure}

	The present model directly distinguishes the sliding or attachment of the colliding particle, thereby enabling us to observe the particle motions near the bubble surface. Figs.~\ref{fig:visual_075} and~\ref{fig:visual_12} show the particle trajectories near the bubble. The bubbles are represented using the iso-surfaces of the volume fraction $c$ of 0.5.  
	
	The particle trajectories are compared with and without the bubble--particle interaction model. Without the model, no particles interact with the bubble, as shown in Figs.~\ref{fig:visual_075}(a) and~\ref{fig:visual_12}(a). No particles are eliminated even when particles are in contact with the bubble surface. {Other} cases show {that} attached particles are eliminated.
	As previously discussed in Section~\ref{sec:numerical_determination}, attached particles are eliminated on the effective bubble before they are in contact with the actual bubble surface. 
	When the bubble diameter is 0.75 mm, the number of attached particles are almost identical as shown in Figs.~\ref{fig:visual_075}(b) and~\ref{fig:visual_075}(c). {The reason is that} both cases have similar attachment diameters of 0.1 mm (Figs.~\ref{fig:attachment_dia}(a) and~\ref{fig:attachment_dia}(b)). 
	Similarly, Fig.~\ref{fig:visual_12} shows the case of the bubble diameter of 1.2 mm. The number of attached particles of the case {in} Fig.~\ref{fig:visual_12}(b) is larger than that {in} Fig.~\ref{fig:visual_12}(c). The reason is that the attachment diameter of the former case is 90 $\upmu$m, which is larger than the latter case of 60 $\upmu$m (Figs.~\ref{fig:attachment_dia}(c) and~\ref{fig:attachment_dia}(d)).
	
	\subsection{Rising of multiple bubbles with particles}\label{sec:multiple}

\begin{figure}
	\centering
	\includegraphics[width=0.55\textwidth]{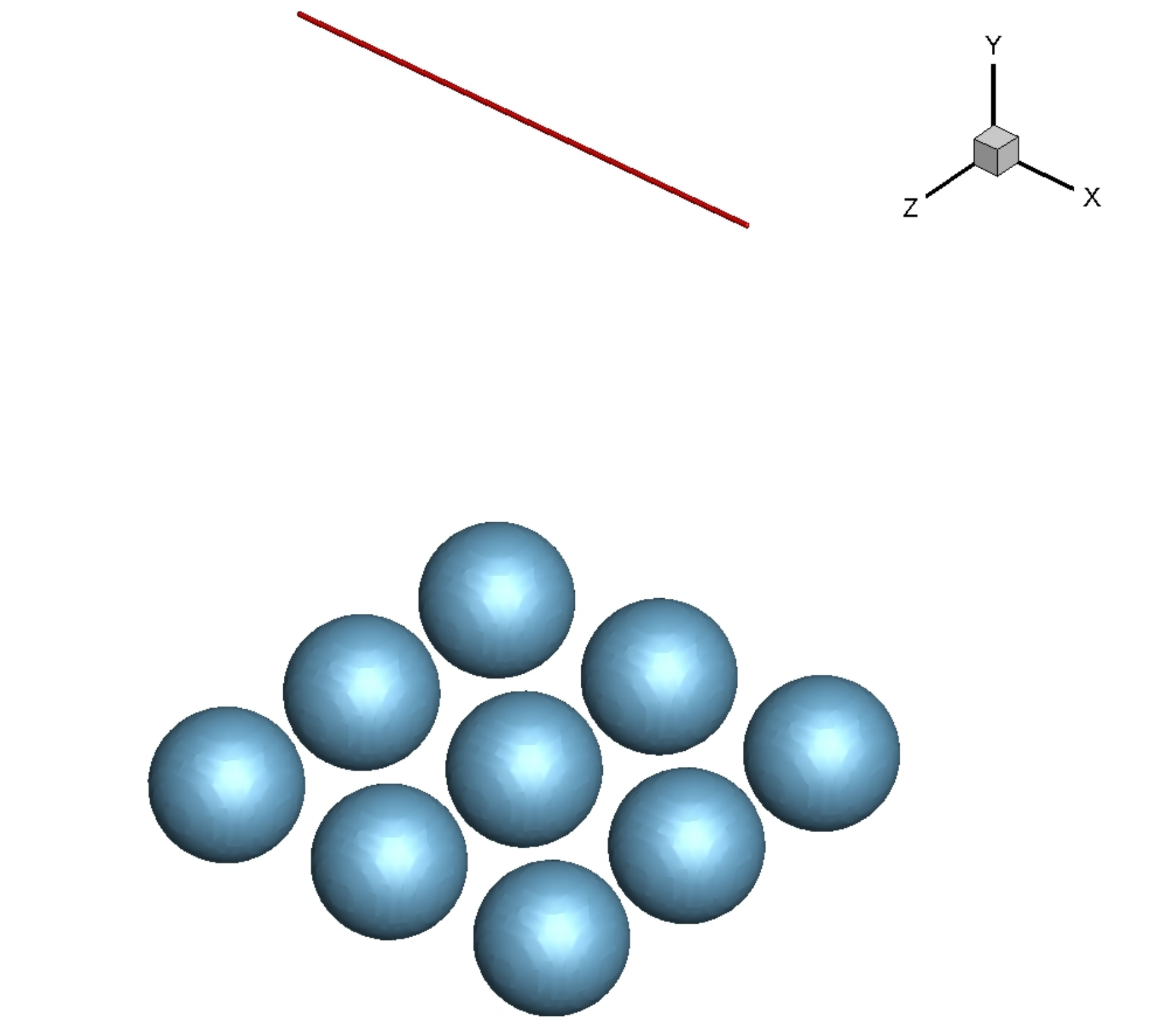}
	\caption{Lateral view of the initial state of the rising of multiple bubbles with particles for $d_\textup{b}=0.75$ mm.}
	\label{fig:multi_075_init}
\end{figure}

\begin{figure}
	\centering
	\includegraphics[width=0.75\textwidth]{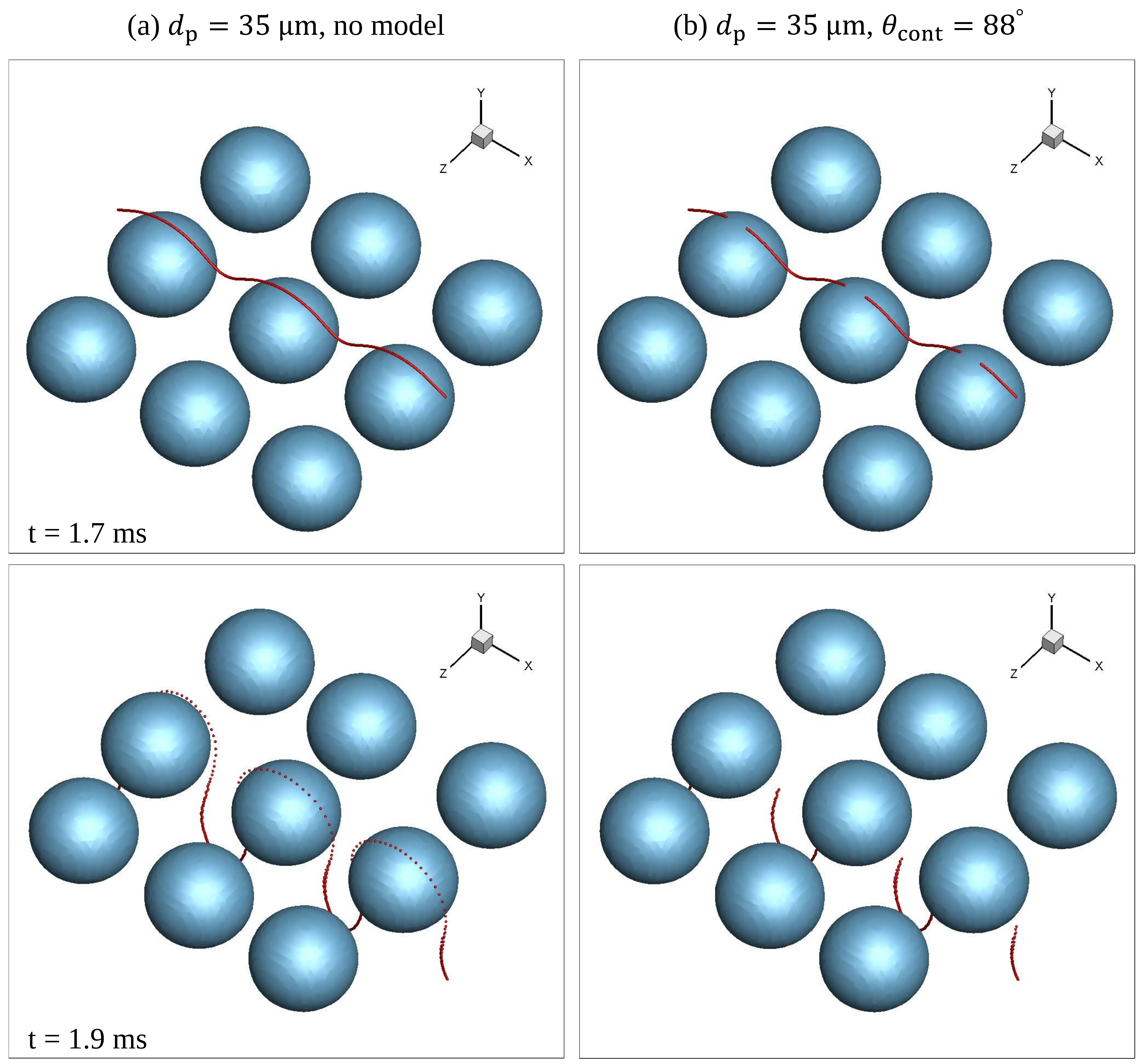}
	\caption{Lateral view of the rising of multiple bubbles with particles for $d_\textup{b}=0.75$ mm and $d_\textup{p}=35$ $\upmu$m (a) without the model and (b) with the model and $\theta_\textup{cont}=88^{\circ}$ at the same time and location.}
	\label{fig:multi_075_3d}
\end{figure}

\begin{figure}[t]
	\centering
	\includegraphics[width=0.6\textwidth]{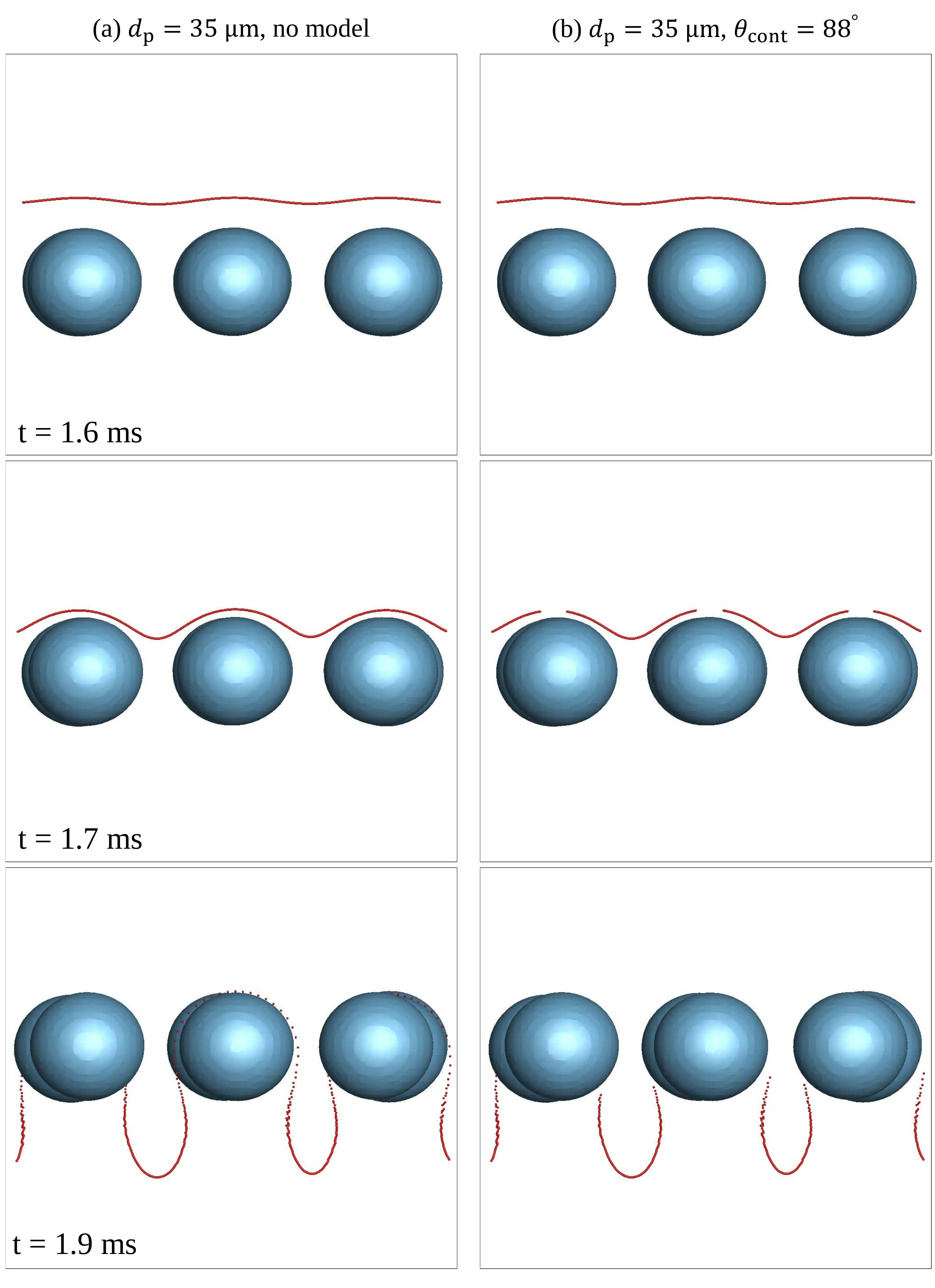}
	\caption{Side view ($xy$-plane) of the rising of multiple bubbles with particles for $d_\textup{b}=0.75$ mm and $d_\textup{p}=35$ $\upmu$m (a) without the model and (b) with the model and $\theta_\textup{cont}=88^{\circ}$ at the same time and location.}
	\label{fig:multi_075_2d}
\end{figure}
	The present bubble--particle interaction model can be applied to multiple bubbles. 
	The computational domain is equivalent to the single bubble case, as presented in Section~\ref{sec:single}. Periodic boundary conditions are applied to the side walls instead of using the slip wall conditions. Fig.~\ref{fig:multi_075_init} shows {that} the $3 \times 3$ stationary spherical bubbles with the diameter of 0.75 mm are initialized at the equal distance from the bottom wall. The particles with the diameter of 35 $\upmu$m are uniformly distributed above the bubbles along the $x$-axis in the middle of the $yz$-plane. Two cases are simulated with and without the bubble--particle interaction model.

	\subsubsection{Particle trajectory}\label{sec:visualization_multiple}
		
	For multiple bubbles, lateral and side views are presented to visualize the particle trajectories near the multiple bubbles. Figs.~\ref{fig:multi_075_3d} and~\ref{fig:multi_075_2d} show that the lateral view and the side view of the rising of multiple bubbles with particles. No particles are attached to the bubble surface without the interaction model, as shown in Figs.~\ref{fig:multi_075_3d}(a) and~\ref{fig:multi_075_2d}(a). By contrast, Figs.~\ref{fig:multi_075_3d}(b) and~\ref{fig:multi_075_2d}(b) show that the attached particles are eliminated on the effective bubble surface. 
	This case shows that the present bubble--particle interaction model can predict particle motions on the bubble surface regardless of the number of bubbles.
	
	
	\section{Conclusions}\label{sec:conclusions}

	Numerical algorithms for simulating gas--liquid--solid flows, including bubble--particle interaction, have been proposed . Phase interfaces are described and advected by the geometrical conserving VOF method. The kinematics of the particles is solved in the Lagrangian framework. The bubble--particle interaction {is} modeled in the VOF framework to predict the particle motions on the bubble surface, including collision, sliding, and attachment.
	The new algorithms {consist} of two procedures, namely, detecting collision and determining sliding or attachment for the colliding particle. 
	These algorithms require the angle between two vectors, the bubble velocity, and the vector of the particle position from the bubble center. The calculation of this angle is proposed using the geometric VOF method.
	In addition, the use of the effective bubble is suggested because the exact point of collision cannot be obtained in the VOF framework. 
	
	Suggested algorithms are validated by reproducing experimental cases. The rising of the single bubble with particles is simulated in terms of bubble diameters, particle diameters, and contact angles. 
	The present numerical model successfully predicts the collision and attachment probabilities.
	Moreover, particle trajectories show how particles pass by the bubble with and without the present model. We can observe that attached particles are eliminated. The cases of rising of multiple bubbles indicate that the present method works irrespective of the number of bubbles.

	\section*{Acknowledgements}
	This research was supported by the National Research Foundation of Korea (NRF) under the Project Number NRF-2017R1E1A1A033070514, the Korea Institute of Energy Technology Evaluation and Planning (KETEP), the Ministry of Trade, Industry $\&$ Energy (MOTIE) of the Republic of Korea (No. 20193020020160), and POSCO under  Project Number 2018Y077.
	
	
	
	\appendix
	\section{Tuning the effective collision angle and critical angle}\label{sec:tunning}
	
	Numerical collision is defined on the basis of the effective bubble surface. The effective collision angle $\theta_\textup{m}^{\ast}$ and the critical angle $\theta_\textup{cr}^{\ast}$ {must} be modeled from the actual collision angle $\theta_\textup{m}$ and the critical angle $\theta_\textup{cr}$. Eqs~(\ref{eq:tunning_collision}) and~(\ref{eq:tunning_critical}) are tuning equations. Parametric studies show that the effective factors $k_\textup{m}$ and $k_\textup{cr}$ have linear relations for the particle diameter $d_\textup{p}$. Thus, the effective factors $k_\textup{m}$ and $k_\textup{cr}$ are written as follows:
	\begin{subequations}\label{eq:tunning1}
	\begin{align}
	&k_\textup{m}=f_\textup{m} \times d_\textup{p} + g_\textup{m}, \\
	&k_\textup{cr}=f_\textup{cr} \times d_\textup{p} + g_\textup{cr}, 
	\end{align}
	\end{subequations}
where $f_\textup{m}$ and $g_\textup{m}$ are {the} functions of the bubble diameter $d_\textup{b}$; and $f_\textup{cr}$ and $g_\textup{cr}$ are {the} functions of the bubble diameter $d_\textup{b}$ and the contact angle $\theta_\textup{cont}$, respectably:
	\begin{subequations}\label{eq:tunning2}
	\begin{align}
	&f_\textup{m} = a_1 \times d_\textup{b} + a_2, \label{eq:f_m}\\
	&g_\textup{m} = a_3 \times d_\textup{b} + a_4, \label{eq:g_m}\\
	&f_\textup{cr} = b_1 \times d_\textup{b} + b_2 \times \sin(\theta_\textup{cont}/2) + b_3 \times d_\textup{b} \times \sin(\theta_\textup{cont}/2) + b_4, \label{eq:f_cr}\\
	&g_\textup{cr} = b_5 \times d_\textup{b} + b_6 \times \sin(\theta_\textup{cont}/2) + b_7 \times d_\textup{b} \times \sin(\theta_\textup{cont}/2) + b_8, \label{eq:g_cr}
	\end{align}
	\end{subequations}
where coefficients $a_i$ for $i=[1,4]$ and $b_i$ for $i=[1,8]$ are the constants to be determined. The contact angle $\theta_\textup{cont}$ is expressed as a sinusoidal form similar to Eq.~(\ref{eq:attachment_probability}). The collision angle $\theta_\textup{m}$ is not a function of the contact angle $\theta_\textup{cont}$ as shown in Eqs.~(\ref{eq:collision_diameter}) and~(\ref{eq:collision_probability}). Therefore, $f_\textup{m}$ and $g_\textup{m}$ are {only the} functions of the bubble diameter $d_\textup{b}$.

	The coefficients in Eqs.~(\ref{eq:tunning1}) and~(\ref{eq:tunning2}) are determined by the following procedures. First, the effective factors $k_\textup{m}$ and $k_\textup{cr}$ need to be determined. The values are adjusted until the collision and attachment probability obtained from the simulation match with Eqs.~(\ref{eq:collision_probability}) and~(\ref{eq:attachment_probability}). Then, the effective factors are linearly fitted; thus, $f_\textup{m}$, $f_\textup{cr}$, $g_\textup{m}$, and $g_\textup{cr}$ are obtained. Coefficients $a_i$ and $b_i$ can be determined by substituting the bubble diameters and the contact angles into Eqs.~(\ref{eq:f_m}) to~(\ref{eq:g_cr}). In this study, the bubble diameters 0.75 mm and 1.2 mm and the contact angles of $50^{\circ}$ and $88^{\circ}$ are used. This tunning procedures need to be repeated, and the coefficients need to be reconstructed if the size of the effective bubble is changed.
	
	

	\bibliographystyle{unsrt}
	\bibliography{bibfile}
	
	


	

	\newpage

\pagebreak
\clearpage

\thispagestyle{empty}
\section*{Significance and novelty}
\begin{itemize}
\item{
A novel numerical framework for simulating gas--liquid--solid flows is proposed with the bubble--particle interaction model.
}
\item{
New algorithms for detecting collision and determining sliding or attachment in a VOF framework are suggested using the geometrical relationship between a bubble and a particle and by adopting the concept of an effective bubble.
}
\item{
The present numerical model for bubble--particle interaction is validated and visualized using simulation cases of rising of a single bubble and rising of multiple bubbles with particles in terms of bubble diameters, particle diameters, and contact angles.
}
\end{itemize}

\newpage
\section*{Highlights}
\begin{itemize}
\item{
Simulating gas--liquid--solid flows by combining an Eulerian approach and a Lagrangian approach
}
\item{
Proposing algorithms for detecting collision and determining sliding or attachment
}
\item{
Validating the numerical method based on the collision and attachment probability
}
\item{
Visualizing bubbles through particles with and without using the bubble--particle interaction model
}
\end{itemize}	

\end{document}